\documentclass[11pt,a4paper]{article}
\pdfoutput=1
\usepackage{jheppub}
\usepackage{graphicx}
\usepackage{lipsum}
\usepackage[all]{hypcap}
\usepackage{multirow}

\newcommand{\sS}{{\cal S}}
\newcommand{\cg}{{c_g}}

\newcommand{\cz}{{c_Z}}
\newcommand{\cw}{{c_W}}
\newcommand{\cwh}{{\widehat{c_W}}}
\newcommand{\czh}{{\widehat{c_Z}}}
\newcommand{\ca}{{c_\gamma}}
\newcommand{\cza}{{c_{Z\gamma}}}
\newcommand{\Gun}{\Gamma_{\text{other}}}

\newcommand{\BRp}{{\rm BR}_p}
\newcommand{\BRu}{{\rm BR}_{\text{other}}}
\newcommand{\BRa}{{\rm BR}_{\gamma\gamma}}

\newcommand{\bss}{\begin{tiny}}
\newcommand{\ess}{\end{tiny}}
\newcommand{\eq}[1]{eq.~(\ref{#1})}

\newcommand{\nn}{\nonumber}

\newcommand{\be}{\begin{equation}}
\newcommand{\ee}{\end{equation}}
\newcommand{\bea}{\begin{eqnarray}}
\newcommand{\eea}{\end{eqnarray}}
\newcommand{\bc}{\begin{center}}
\newcommand{\ec}{\end{center}}

\begin{document}

\title{Interpreting a 750 GeV Diphoton Resonance}

\affiliation[a]{Department of Particle Physics and Astrophysics,
Weizmann Institute of Science,\\ Rehovot 7610001, Israel}
\affiliation[b]{Department of Physics and Astronomy,
University of Sussex,\\ Brighton, BN1 9QH, UK \vspace{2mm}}

\author[a]{Rick S. Gupta,}
\author[a,b]{Sebastian J\"{a}ger,}
\author[a]{Yevgeny Kats,}
\author[a]{Gilad Perez,}
\author[a]{Emmanuel Stamou}

\emailAdd{rsgupta@weizmann.ac.il}
\emailAdd{S.Jaeger@sussex.ac.uk}
\emailAdd{yevgeny.kats@weizmann.ac.il}
\emailAdd{gilad.perez@weizmann.ac.il}
\emailAdd{emmanuel.stamou@weizmann.ac.il}

\abstract{
We discuss the implications of the significant excesses in the
diphoton final state observed by the LHC experiments
ATLAS and CMS around a diphoton invariant mass of $750$ GeV.
The interpretation of the excess  as a spin-zero  
$s$-channel resonance implies model-independent lower bounds
on both its branching ratio and its coupling to photons, which
stringently constrain dynamical models.
We consider both the case where the excess is described by a 
narrow and a broad resonance.
We also obtain model-independent constraints on the allowed
couplings and branching fractions to final states other than diphotons, 
by including the interplay with 8 TeV searches. 
These results can guide attempts to construct viable dynamical 
models of the resonance.
Turning to specific models, our findings suggest that the anomaly cannot 
be accounted for by the presence of only an additional singlet or doublet 
spin-zero field and the Standard Model degrees of freedom; this includes 
all two-Higgs-doublet models.
Likewise, heavy scalars in the MSSM cannot explain the excess
if stability of the electroweak vacuum is required, at least in a leading-order analysis.
If we assume that the resonance is broad we find that it is
challenging to find a weakly coupled explanation.
However, we provide an existence proof in the form of a model with vectorlike quarks with large
electric charge that is perturbative up to the 100~TeV scale.
For the  narrow-resonance case a similar model can be 
perturbative up to high scales also with smaller charges. 
We also find that, in their simplest form,  dilaton models  cannot explain the size of the excess.
Some implications for flavor physics are briefly discussed.}

\maketitle

\section{Introduction}
Very recently, both the ATLAS and the CMS collaborations at CERN have reported
mutually consistent ``bumps'' in the diphoton invariant mass distribution 
around $750$~GeV~\cite{ATLAS-CONF-2015-081,CMS-PAS-EXO-15-004}. 
Based on $3.2$ and $2.6$~fb$^{-1}$ of the $13$~TeV LHC data, the corresponding deviations from the 
background-only hypothesis have a local significance of $3.9\sigma$ and $2.6\sigma$ 
in ATLAS and CMS, respectively. 
The bumps are best described by a relative width $\Gamma/M \approx 6\%$ in 
ATLAS~\cite{ATLAS-CONF-2015-081} but a sub-resolution width in 
CMS~\cite{CMS-PAS-EXO-15-004}.
However, this discrepancy is not statistically significant 
and we will generally present results as a function of the unknown width.
The resonant excesses are suggestive of the decay of a new particle 
beyond the Standard Model (BSM).
The kinematic properties of the events in the excess region are reported not to show 
significant deviations compared with events in sidebands. This disfavors significant 
contributions to the production from decays of yet heavier particles or associated production
and motivates focusing on the case of a single production of a $750$~GeV resonance.

The purpose of the present
paper is to characterise this theoretically unexpected result and discuss its
implications for some leading paradigms for physics beyond the
Standard Model. 
It is divided into two main parts, the first of which
comprises a model-independent framework that aims to equip the reader
with handy formulas for interpreting both the signal and the most
important resonance search constraints from existing LHC searches in
the context of BSM models. We derive a number of bounds, including
model-independent lower bounds on the branching ratio and partial
width into photons of the hypothetical new particle. 
The second part investigates concrete scenarios, including the 
possibility of interpreting the resonance as
the dilaton in a theory with spontaneous breaking of scale invariance or as a heavy Higgs
scalar a two-Higgs-doublet model (2HDM). 
We find the properties of the observed excess to be quite constraining. 
In particular, a leading-order analysis suggests that the interpretation as
an $s$-channel resonance, if confirmed,
cannot be accommodated within the Minimal Supersymmetric Standard Model (MSSM)
even under the most conservative assumptions about the MSSM parameters
and the true width of the resonance; this conclusion holds if we
require  the absence of charge- and colour-breaking minima.

\section{Model-independent constraints}
\label{sec:model-independent}

We start by discussing what can be inferred about the new hypothetical particle from data alone. 
We will first describe the implications of the observed properties of the diphoton bumps, 
and then examine the constraints from the absence of significant excesses in resonance 
searches in other final states that could be sensitive to other decay modes of the same particle.

\subsection{Implications of the excess alone}
\label{sec:lag}

Both ATLAS and CMS observe excesses in a diphoton invariant mass region
near 750~GeV~\cite{ATLAS-CONF-2015-081,CMS-PAS-EXO-15-004}. 
For the purposes of this work, we will generally assume the signal 
contribution to be $N=20$ events for ${\cal L}=5.8~{\rm fb}^{-1}$ integrated 
luminosity (adding up ATLAS and CMS), but will make clear the scaling of our findings 
with $N$ wherever feasible. 
We will assume a signal efficiency (including acceptance) of $\epsilon = 50\%$, 
even though, in general, this does have some dependence on both the experiment 
and the details of the signal model.

The most straightforward signal interpretation is resonant
$s$-channel production of a new unstable particle.
The observed signal strength corresponds to a 13\,TeV inclusive
cross section to diphotons of
\begin{equation}
\sigma_{13}\times\BRa \approx 6.9~\mbox{fb} \times \left(\frac{N}{20}
\right) \left( \frac{50\%}{\epsilon} \right)
\left(\frac{5.8~{\rm fb}^{-1}}{{\cal L}_{13}} \right)\,.
\label{sigmaBR13}
\end{equation}

The diphoton final state precludes a spin-one interpretation due to the
Landau-Yang theorem~\cite{Landau:1948kw,Yang:1950rg}, and we will henceforth assume 
spin zero.
We take the mass to be $M=750$\,GeV; small variations have no
significant impact on our findings. The shape of the excess in ATLAS may indicate
a width of about $\Gamma=45$\,GeV~\cite{ATLAS-CONF-2015-081}. 
However, we will also contemplate the case of smaller width below, 
and discuss how our main findings depend on this.

A minimal dynamical input is necessary to interpret the result and 
incorporate $8$\,TeV LHC constraints.
The width-to-mass ratio is small enough to justify a narrow-width
approximation to the level of accuracy we aim at here. 
In the narrow-width limit, resonant scattering amplitudes factorize into
production and decay vertices, which we parameterize by terms in a 
``Lagrangian'' for the resonance $\sS$,
\begin{multline}
{\cal L} = 
           -\frac{1}{16\pi^2} \frac{1}{4} \frac{\ca}{M} \sS F^{\mu\nu}F_{\mu\nu} 
           -\frac{1}{16\pi^2} \frac{1}{4} \frac{\cg}{M} \sS G^{\mu\nu,a}G^a_{\mu\nu}\\ 
           -\frac{1}{16\pi^2} \frac{1}{2} \frac{\cw}{M} \sS W^{-\mu\nu}W^+_{\mu\nu} 
           -\frac{1}{16\pi^2} \frac{1}{4} \frac{\cz}{M} \sS Z^{\mu\nu}Z_{\mu\nu} 
           -\frac{1}{16\pi^2} \frac{1}{4} \frac{\cza}{M} \sS Z^{\mu\nu}F_{\mu\nu} \\
           - \cwh{m_W} \sS W^{+\,\mu}W^-_{\mu} 
           -\frac{1}{2} \czh{m_Z} \sS Z^{\mu}Z_{\mu} 
	   - \sum_{f} c_f \sS \bar f f\,.
\label{eq:lagrangian}
\end{multline}
In this parametrization, $M$ is the mass of the resonance $\sS$.
We emphasize that each term denotes a particular production and/or
decay vertex  and that the parameterization ${\cal L}$ does
not make any assumptions about hierarchies of scales.\footnote{In particular,
the ``couplings'' $c_i$ are on-shell form factors that generally
include contributions from light particles and CP-even phases due to
unitarity cuts. Contributions from particles with mass
$\gg M$ can be matched to a local effective Lagrangian similar to
eq.~\eqref{eq:lagrangian}.
We discuss examples in section~\ref{sec:models}.}
If ${\cal S}$ is a pseudoscalar, $\cwh=\czh=0$, while all the other couplings
lead to the same results as we present in this section for the scalar upon the
replacements $\sS \bar f f \to i \sS \bar f \gamma^5 f$, $X^{\mu\nu}X'_{\mu\nu} \to X^{\mu\nu}\tilde{X}'_{\mu\nu}$, 
where $X^{(\prime)\mu\nu} = F^{\mu\nu}$, $G^{\mu\nu,a}$, $W^{\pm\mu\nu}$, $Z^{\mu\nu}$ 
(up to minor differences in the phase-space factors from table~\ref{tab:partialwidth} below).

The total decay width of $\sS$ imposes one constraint on the couplings,
\begin{equation}\label{eq:width}
\frac{\Gamma}{M} =  \sum_{i} \frac{\Gamma_i}{M} =
\sum_i n_i |c_i|^2 \approx 0.06 \,,
\end{equation}
where the (dimensionless) coefficients $n_i$ are listed 
in table~\ref{tab:partialwidth} for the modes considered in the present analysis.
In particular,  eq.~\eqref{eq:width} directly
implies upper bounds on the magnitude of each $c_i$,
since observations imply that the width cannot exceed the ATLAS-preferred value of 45~GeV by more than a factor of about two.

\begin{table}[t]
\begin{center}
\begin{tabular}{ccc}
mode & Width coefficient $n_i$ & $n_i$ (\#)\\\hline\hline
$\gamma\gamma$ 	&$ \frac{1}{16 (4\pi)^5}$ 					& $1.99 \times 10^{-7}$		\\
$gg$ 		& $\frac{1}{2 (4\pi)^5}$  					& $1.60 \times 10^{-6}$		\\[0.5em]
$q_i \bar q_i$ 	& $\frac{3}{8\pi}$ 						& $0.119$	\\[0.5em]
%$t \bar t$ 	& $\frac{3}{8\pi} (1 - 4 m_t^2/M^2)^{3/2}$ 			& $0.0834$ 	\\[0.5em]
$\widehat{WW} $ 	& $\frac{1}{64\pi} \sqrt{1 - 4 m_W^2/M^2} \frac{M^2}{m_W^2}
\left(1-4\frac{m_W^2}{M^2}+12\frac{m_W^4}{M^4}\right)$				& $0.404$  	\\[0.5em]
$\widehat{ZZ} $ 		& $\frac{1}{128\pi} \sqrt{1 - 4 m_Z^2/M^2}\frac{M^2}{m_Z^2}
\left(1-4\frac{m_Z^2}{M^2}+12\frac{m_Z^4}{M^4}\right)$   		    	& $0.154$ 	\\[1em]
$WW$ 	& $\frac{1}{8 (4\pi)^5} \sqrt{1 - 4 m_W^2/M^2} 
\left(1-4\frac{m_W^2}{M^2}+6\frac{m_W^4}{M^4}\right)$				& $3.72\times 10^{-7}$\\[0.5em]
$ZZ$ 		& $\frac{1}{16 (4 \pi)^5} \sqrt{1 - 4 m_Z^2/M^2}
\left(1-4\frac{m_Z^2}{M^2}+6\frac{m_Z^4}{M^4}\right)$   		    	& $1.82\times 10^{-7}$\\[0.5em]
$Z\gamma$ 	& $\frac{1}{32 (4 \pi)^5} 
\left(1-\frac{m_Z^2}{M^2}\right)^3$   		    				& $9.54\times 10^{-8}$\\[1em]
\hline
\end{tabular}
\end{center}
\caption{Width coefficients.\label{tab:partialwidth}}
\end{table}

It is possible and convenient to represent the observed signal in terms of the branching ratios
to the production mode and to $\gamma\gamma$.
If a single production mode, $p$, dominates, the number of signal events, $N$, in the  
$13$\,TeV analyses fixes the product
\begin{equation}
\BRa\times\BRp = n_p \frac{M}{\Gamma}\frac{N}{ 
\epsilon x_\sS^{13,p}
{\cal L}_{13}
}
= \kappa_p\times
			\left(\frac{N}{20}\right)%
			\left(\frac{45~{\rm GeV}}{\Gamma}\right)
                        \left(\frac{5.8~{\rm fb}^{-1}}{{\cal L}_{13}} \right) \,,
\label{eq:BRprod}
\end{equation}
where,  for the production modes mediated by the various couplings from eq.~\eqref{eq:lagrangian},
\begin{equation}
\kappa_p \approx
\{
\overset{gg            }{\rule{0pt}{1.45em}\smash{2.5}},\, 
\overset{u\bar u       }{\rule{0pt}{1.5em}\smash{5.5}},\, 
\overset{d\bar d       }{\rule{0pt}{1.5em}\smash{8.9}},\, 
\overset{s\bar s       }{\rule{0pt}{1.5em}\smash{96}},\, 
\overset{c\bar c       }{\rule{0pt}{1.5em}\smash{140}},\, 
\overset{b\bar b       }{\rule{0pt}{1.5em}\smash{310}},\, 
\overset{WW            }{\rule{0pt}{1.5em}\smash{1600}},\, 
\overset{\widehat{WW}}{\rule{0pt}{1.5em}\smash{16000}},\, 
\overset{ZZ            }{\rule{0pt}{1.5em}\smash{2400}},\, 
\overset{\widehat{ZZ}}{\rule{0pt}{1.5em}\smash{21000}},\, 
\overset{Z\gamma       }{\rule{0pt}{1.45em}\smash{1400}},\, 
\overset{\gamma\gamma  }{\rule{0pt}{1.45em}\smash{170}} 
\}\times 10^{-5} .
\end{equation}
We used the leading-order $\sqrt s = 13$~TeV production cross sections for $M = 750$~GeV,
\be
\sigma_{13} = |c_p|^2 x_\sS^{13,p} \,,
\ee
where $x_\sS^{13,p}$ are listed in table~\ref{tab:ppaa}.\footnote{Results for VBF production, here and below, involve the use of the $\sS WW$ and $\sS ZZ$ vertices in eq.~\eqref{eq:lagrangian} implemented in MadGraph~\cite{Alwall:2014hca} using FeynRules~\cite{Alloul:2013bka}. This is correct in either of the following two situations: (i) the origin of the vertices is local physics, originating in scales $\gg M$, such as in the dilaton case in section~\ref{sec:dilaton}; in such a case the vertices can be interpreted as a unitary-gauge Lagrangian couplings and be used off shell; or (ii) the production process is dominated by nearly on-shell $W$, $Z$ bosons (the same prerequisite under which the equivalent-boson approximation~\cite{Dawson:1984gx,Kane:1984bb} is justified). }
A direct consequence of eq.~\eqref{eq:BRprod}  is a lower bound on the branching ratio 
into photons,
\begin{align}
  \BRa > \kappa_p \left(\frac{N}{20}\right)%
			\left(\frac{45~{\rm GeV}}{\Gamma}\right)
                        \left(\frac{5.8~{\rm fb}^{-1}}{{\cal L}_{13}} \right) .
\label{BRgmgmMin}
\end{align}
Note that this bound becomes more stringent if the width of the
resonance is reduced.

\begin{table}[t]
\centering
\begin{tabular}{cccccccc}
$\sqrt{s}$ 	& [pb] 		
&$gg$ 			& $u\bar u$	& $d\bar d$	& $s\bar s$	& $c\bar c$	& $b\bar b$ \\\hline\hline\\[-1em]
13              &$x^{13,p}_\sS$    %& {\tt NNPD23LO}
&$7.5\cdot 10^{-3}$	&$250$ &$150$ &$14$ &$9.8$ &$4.4$\\
8               &$x^{8,p}_\sS$     %& {\tt NNPD23LO}
&$1.7\cdot 10^{-3}$ &$95$  &$57   $ &$3.7$ &$2.3$ &$0.96$\\\hline
13/8            & $r_p$ %& {\tt NNPDF23LO}
&$4.4$ 			&$2.6$ &$2.7$ &$3.9$ &$4.2$ &$4.6$\\
\hline
\end{tabular}\\[1em]

\begin{tabular}{ccccccccc}
$\sqrt{s}$ 	 & [pb]	&${WW}$			&${\widehat{WW}}$	&${ZZ}$			&$\widehat{ZZ}$		&${Z\gamma}$		&$\gamma\gamma$ \\\hline\hline\\[-1em]
13   &$x^{13,p}_\sS$    &$2.7\cdot 10^{-6}$	&$0.30$			&$8.7\cdot 10^{-7}$	&$8.3\cdot 10^{-2}$	&$7.8\cdot 10^{-7}$	&$1.4 \cdot 10^{-5}$\\
8    &$x^{8,p}_\sS$    	&$6.5\cdot 10^{-7}$	&$6.9\cdot 10^{-2}$ 	&$2.1\cdot 10^{-7}$	&$1.9\cdot 10^{-2}$	&$2.1\cdot 10^{-7}$ 	&$4.7 \cdot 10^{-6}$\\\hline
13/8 &$r_p$ 		&$4.1$ 			&$4.3$ 			&$4.2$ 			&$4.2$ 			&$3.7$ 	&$2.9$\\\hline
\end{tabular}
\caption{Leading-order production cross sections for a resonance with $M = 750$~GeV for couplings
$c_p=1$, at the 13~TeV and 8~TeV LHC, and their ratio, $r_p$.
We have used the leading-order PDF set {\tt NN23LO1}~\cite{Ball:2012cx}
for the predictions of production via $gg$, $q\bar q$, $WW$, $\widehat{WW}$, $ZZ$ and $\widehat{ZZ}$.
For $Z\gamma$ initiated production we use the {\tt CTEQ14QED} PDF set~\cite{Schmidt:2015zda} with photon PDF, 
while for $\gamma\gamma$ fusion we use the results of ref.~\cite{Harland-Lang:2016qjy},
which also discusses the validity of $\gamma\gamma$ fusion results obtained with various PDF sets.
For the $gg$ and $q\bar q$ modes, the process is $pp\to\sS$.
For $WW$, $\widehat{WW}$, $ZZ$ and $\widehat{ZZ}$, both VBF ($pp\to\sS+jj$)
and associated production ($pp\to\sS+W/Z$) contribute.
The latter is small for $\widehat{WW}$ and $\widehat{ZZ}$ 
(approximately $1\%$ of the inclusive value), but is significant for production via
the field-strength $WW$ and $ZZ$ operators (approximately $20\%$ and $30\%$ for  $WW$
and $ZZ$, respectively, at 13 TeV; see also ref.~\cite{Altmannshofer:2015xfo}).
Finally, for production via $Z\gamma$, the processes
$pp\to\sS jj$, $pp\to\sS Z$, and $pp\to\sS \gamma$ contribute with relative weights $94\%$,
$2.6\%$, and $3.6\%$, respectively.
\label{tab:ppaa}}
\end{table}

Alternatively, the excess events fix the product of couplings
\begin{equation}
|\ca c_p| = \sqrt{n_\gamma^{-1} \frac{\Gamma}{M}\frac{N}{ 
\epsilon x_\sS^{13,p}
{\cal L}_{13}}}
= \rho_p\times \sqrt{ 
			\left(\frac{N}{20}\right)
			\left(\frac{\Gamma}{45~{\rm GeV}}\right)
			\left(\frac{5.8~{\rm fb}^{-1}}{{\cal L}_{13}}\right)} \;,
\label{eq:cacp}
\end{equation}
where
\begin{equation}
\rho_p \approx 
\{
\overset{gg            }{\rule{0pt}{1.45em}\smash{530}},\, 
\overset{u\bar u       }{\rule{0pt}{1.5em}\smash{2.9}},\, 
\overset{d\bar d       }{\rule{0pt}{1.5em}\smash{3.7}},\, 
\overset{s\bar s       }{\rule{0pt}{1.5em}\smash{12}},\, 
\overset{c\bar c       }{\rule{0pt}{1.5em}\smash{15}},\, 
\overset{b\bar b       }{\rule{0pt}{1.5em}\smash{22}},\, 
\overset{WW            }{\rule{0pt}{1.5em}\smash{28000}},\, 
\overset{\widehat{WW}}{\rule{0pt}{1.5em}\smash{84}},\, 
\overset{ZZ            }{\rule{0pt}{1.5em}\smash{49000}},\, 
\overset{\widehat{ZZ}}{\rule{0pt}{1.5em}\smash{160}},\, 
\overset{Z\gamma       }{\rule{0pt}{1.45em}\smash{52000}},\, 
\overset{\gamma\gamma  }{\rule{0pt}{1.45em}\smash{12000}} 
\}\,.
\label{eq:rhos}
\end{equation}

Importantly, increasing the production couplings, $c_p$, increases
also the decay rates to the production modes. Since these compete with
the $\gamma\gamma$ decay, $\ca$ cannot be arbitrarily small.
The smallest possible $|\ca|$ corresponds to the situation where the total width is dominated 
by the production mode (which in particular implies  $\Gamma_{\gamma\gamma} \ll \Gamma_p$, for production modes other than $\gamma\gamma$). 
Since the dependence on $|c_p|^2$ cancels
between the production cross section and the diphoton branching
fraction in this limit, this bound is independent of $\Gamma$. We hence have the following model-independent lower bounds on $c_\gamma$:
\begin{multline}
|\ca| > \sqrt{\frac{n_p}{n_\gamma}\,\frac{N}{\epsilon x_\sS^{13,p}{\cal L}_{13}}}=
\{
\overset{gg            }{\rule{0pt}{1.45em}\smash{2.7}},\, 
\overset{u\bar u       }{\rule{0pt}{1.5em}\smash{4.1}},\, 
\overset{d\bar d       }{\rule{0pt}{1.5em}\smash{5.2}},\, 
\overset{s\bar s       }{\rule{0pt}{1.5em}\smash{17}},\, 
\overset{c\bar c       }{\rule{0pt}{1.5em}\smash{21}},\, 
\overset{b\bar b       }{\rule{0pt}{1.5em}\smash{31}},\, 
\overset{WW            }{\rule{0pt}{1.5em}\smash{70}},\, 
\overset{\widehat{WW}}{\rule{0pt}{1.5em}\smash{220}},\, 
\overset{ZZ            }{\rule{0pt}{1.5em}\smash{85}},\, 
\overset{\widehat{ZZ}}{\rule{0pt}{1.5em}\smash{250}},\, 
\overset{Z\gamma       }{\rule{0pt}{1.45em}\smash{65}},\, 
\overset{\gamma\gamma  }{\rule{0pt}{1.5em}\smash{23}}
\}\times\\
\times \sqrt{\left(\frac{N}{20}\right)\left(\frac{5.8~{\rm fb}^{-1}}{{\cal L}_{13}}\right)} \;.
\label{eq:calowerbound}
\end{multline}
If, as it often does, a single production mode dominates in a concrete
model, eq.~\eqref{eq:calowerbound} can be directly  used to identify
how large $\ca$ needs to be.
For production via $\gamma\gamma$ fusion, saturating the
lower bound determines in addition the width to be about 75 MeV.
 In the case where several initial states
contribute, a conservative lower bound is given by
\begin{align}
|\ca| > \sqrt{\frac{n_g}{n_\gamma}\,\frac{N}{\epsilon x_\sS^{13,g}{\cal L}_{13}}}=
2.7
\times \sqrt{\left(\frac{N}{20}\right)\left(\frac{5.8~{\rm fb}^{-1}}{{\cal L}_{13}}\right)} \;.
\label{eq:calowerboundgeneral}
\end{align}
Importantly, eqs.~(\ref{eq:cacp}) and (\ref{eq:rhos}) imply that
  photon fusion accounts for the entire excess once $|c_\gamma| \sim 110$, or less
  if the width is below $45$~GeV. It then follows from eq.~(\ref{eq:calowerbound}) that
  production via the couplings $\widehat{c_W}$ and $\widehat{ c_Z}$ can never be an important
  production mechanism, so we disregard this possibility henceforth. (See
  also the discussion in ref.~\cite{Fichet:2015vvy}.)

\begin{figure}[t]
\begin{center}
\vspace{-0.3cm}
\includegraphics[]{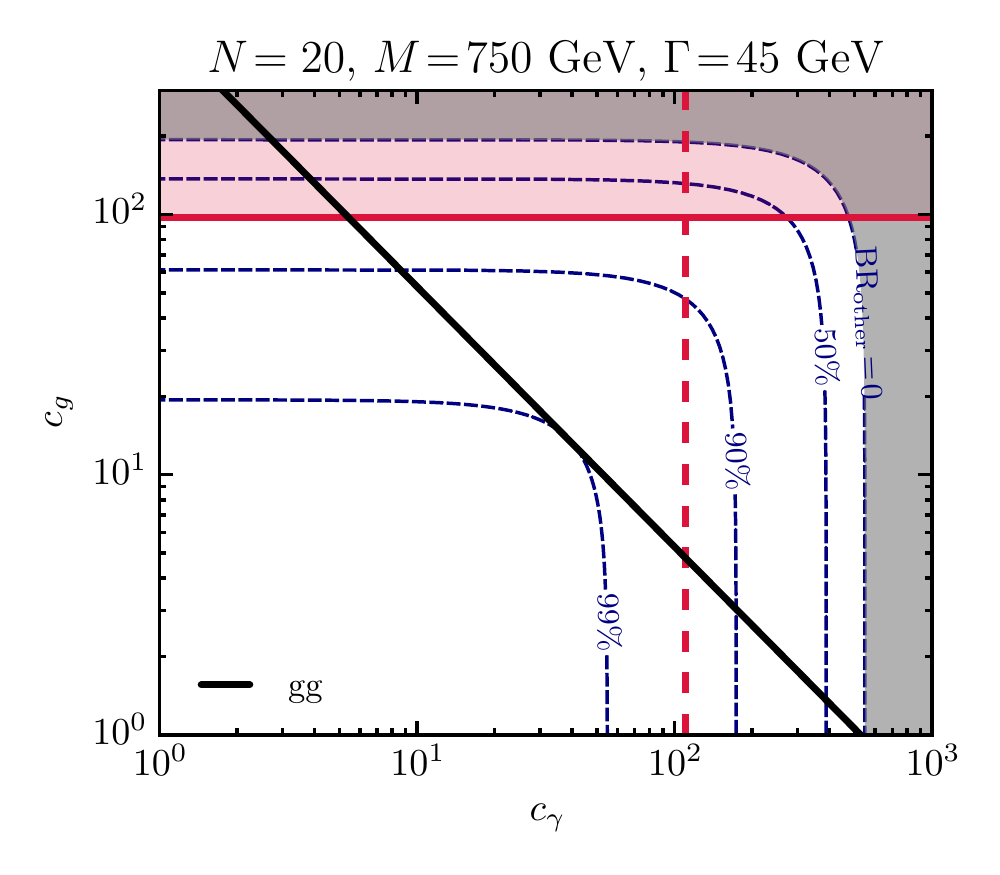}
\vspace{-0.5cm}
\caption{
The black line corresponds to $N = 20$ signal events in the diphoton analyses for $M=750$\,GeV 
and $\Gamma=45$\,GeV when the resonance is produced from $gg$.
Blue dashed lines are contours of fixed branching ratio to modes 
other than $\gamma\gamma$ or $gg$.
The red-shaded area above the thick horizontal line is excluded by dijet resonance 
searches~\cite{CMS:2015neg} due to decays to $gg$ alone.
The shaded gray region corresponds to values of $c_g$, $\ca$ that produce a width
larger than $45$\,GeV.
The vertical dashed red line indicates the $\ca$ value for which photon fusion
alone would account for all signal events even for $\Gamma = 45$~GeV, thus ruling out the region of larger
$\ca$ values.
\label{fig:cacg}}
\end{center}
\end{figure}

\begin{figure}[t]
\begin{center}
\vspace{-0.3cm}
\includegraphics[]{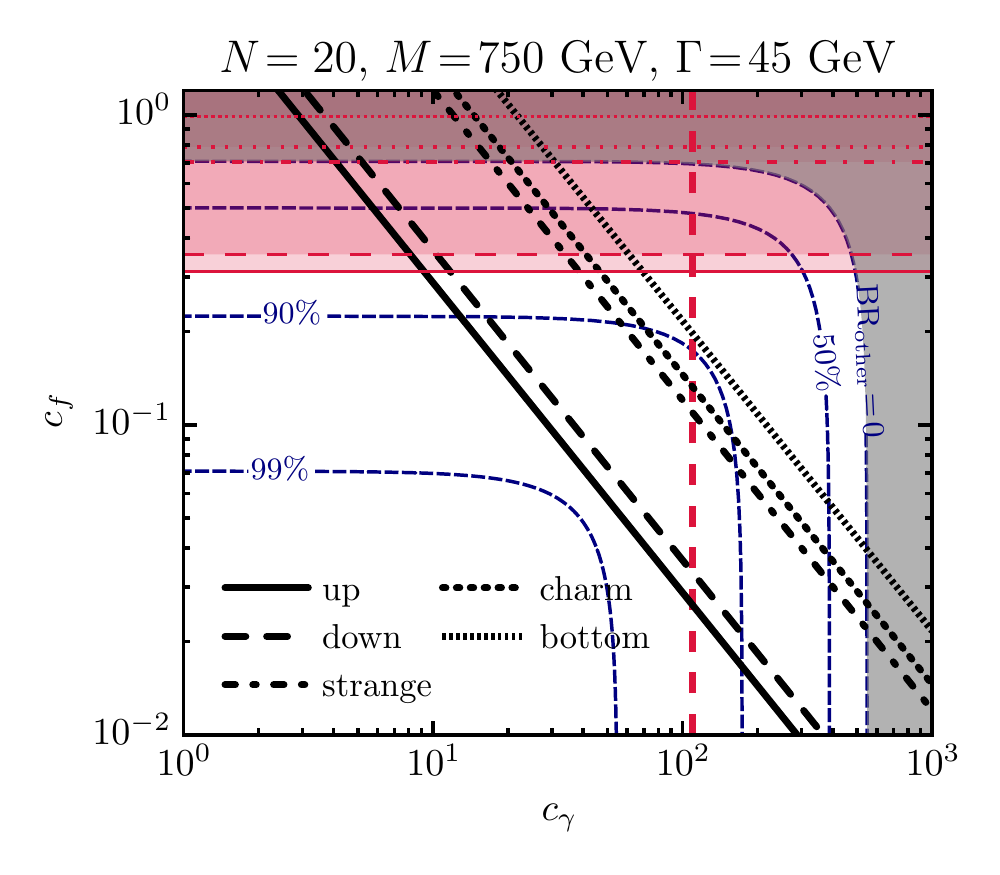}
\vspace{-0.5cm}
\caption{
Black lines correspond to $N = 20$ signal events in the diphoton analyses for $M=750$\,GeV and $\Gamma=45$\,GeV.
Different dashing styles indicate the various production modes,
$u\bar u$, $d\bar d$, $s\bar s$, $c\bar c$, and $b\bar b$.
Blue dashed lines are contours of fixed branching ratio to modes other 
than $\gamma\gamma$ or the production mode.
The red-shaded areas above the various horizontal lines, with dashing 
styles corresponding to the production modes, are excluded by dijet resonance 
searches~\cite{CMS:2015neg} due to decays to the production mode alone. 
The vertical dashed red line indicates the $\ca$ value for which photon fusion
alone would account for all signal events even for $\Gamma = 45$~GeV, thus ruling out the region of larger
$\ca$ values.
\label{fig:cacf}}
\end{center}
\end{figure}

In figures~\ref{fig:cacg} and~\ref{fig:cacf}, we plot the general relation between 
$|\ca|$ and $|c_p|$ for the case of $N = 20$ excess events, switching on 
one production channel at a time.
The mass and total width are fixed at $750$ and $45$\,GeV, respectively. 
The partial widths to diphotons, $\Gamma_{\gamma\gamma}$, and to the production mode, 
$\Gamma_p$, are assumed to be supplemented by decays to other possible final states, 
$\Gun$, to make up the total width:
\be
\Gun \equiv \Gamma - \Gamma_{\gamma\gamma} - \Gamma_p \,.
\ee
Contours of fixed $\BRu\equiv\Gun/\Gamma$ are shown in dashed blue.
From the left panels of the figures it is evident that for a given $\BRu$ there
exist two solutions, one with small and another with large $\ca$.
However, this second solution is generally incompatible
  with the upper limit $|\ca| \lesssim 110$, unless $\BRu$ is close to 100\%.
The gray-shaded regions correspond to values of $c_p$ and $\ca$ for which
the total width is larger than $45$\,GeV. Horizontal red lines and the
corresponding red-shaded regions indicate the parameter space excluded by $8$\,TeV
dijet searches. We discuss them in the next subsection.

\subsection{Interplay with previous LHC searches}

Important additional information about the properties of the new particle 
can be obtained based on the non-observation of any of its decays in 
Run\,1 of the LHC, in particular in the 20~fb$^{-1}$ of data collected at $\sqrt s = 8$\,TeV.

We first  consider limits from the diphoton resonance searches. 
The most relevant limit for the broad resonance hypothesis preferred by the ATLAS excess, 
$\Gamma/M \approx 6\%$, is the CMS 95\% CL limit
\begin{equation}
\sigma_8\times\BRa \lesssim 2.5~\text{fb} \,,
\label{}
\end{equation}
which was derived for scalar resonances with $\Gamma/M = 10\%$~\cite{Khachatryan:2015qba}. For a narrow resonance, which might be preferred by the CMS data, the same search sets the limit
\begin{equation}
\sigma_8\times\BRa \lesssim 1.3~\text{fb} \,.
\label{}
\end{equation}
Somewhat weaker limits, of $2.5$ and $2.0$~fb, were obtained 
by ATLAS~\cite{Aad:2015mna} and CMS~\cite{CMS-PAS-EXO-12-045}, respectively, for RS 
gravitons with $k = 0.1$, which are also narrow.

The compatibility of the observed excesses with the $8$\,TeV diphoton searches depends primarily on the parton luminosity ratio, $r_p$, listed in table~\ref{tab:ppaa},\footnote{More precisely, $r_p$ is the cross sections ratio. For VBF and associated production, it cannot be approximated by the parton luminosity ratio at $\sqrt{\hat s} = 750$~GeV (as was done in some of the recent papers that claimed $r_{\rm VBF} \approx 2.5$) since in most events $\sqrt{\hat s}$ is significantly higher than 750~GeV because of the two forward jets or the additional electroweak boson.}
since the selection efficiencies of the searches are similar. 
The ATLAS+CMS excess, eq.~\eqref{sigmaBR13}, translates to 
\begin{equation}
\sigma_8\times\BRa = \frac{\sigma_{13}\times\BRa}{r_p}
\approx \left(\frac{N}{20}\right)\times
\{
\overset{gg            }{\rule{0pt}{1.45em}\smash{1.6}},\, 
\overset{u\bar u       }{\rule{0pt}{1.5em }\smash{2.6}},\, 
\overset{d\bar d       }{\rule{0pt}{1.5em }\smash{2.6}},\, 
\overset{s\bar s       }{\rule{0pt}{1.5em }\smash{1.8}},\, 
\overset{c\bar c       }{\rule{0pt}{1.5em }\smash{1.7}},\, 
\overset{b\bar b       }{\rule{0pt}{1.5em }\smash{1.5}},\, 
\overset{WW            }{\rule{0pt}{1.5em }\smash{1.7}},\, 
\overset{ZZ            }{\rule{0pt}{1.5em }\smash{1.6}},\, 
\overset{Z\gamma       }{\rule{0pt}{1.45em}\smash{1.8}},\,
\overset{\gamma\gamma  }{\rule{0pt}{1.5em }\smash{2.4}}
\}\,\text{fb}.
\end{equation}
We see that $N = 20$ excess events at $13$~TeV are borderline compatible with the $8$~TeV analyses, especially if the resonance is broad.
The $gg$, heavy-quark and electroweak-boson production modes are somewhat favoured in this respect because their cross sections increase more rapidly with $\sqrt{s}$.
\begin{table}[t]
\begin{center}
\begin{tabular}[]{c|c|ccccccccccc}
\multicolumn{2}{r|}{decay mode $i \to $} & $gg$ & $q\bar q$ & $t\bar t$	& $WW$ & $ZZ$ & $hh$ & $Zh$ & $\tau\tau$ & $Z\gamma$ & $ee+\mu\mu$ \\\hline\hline
\multicolumn{2}{c|}{\multirow{2}{*}{$(\sigma_8\times {\rm BR}_i)^{\rm max}$ [fb]}} & 4000 & 1800 & 500 & 60 & 60 & 50 & 17 & 12 & 8 & 2.4 \\
\multicolumn{2}{c|}{} & \cite{CMS:2015neg} & \cite{CMS:2015neg} & \cite{Khachatryan:2015sma} & \cite{Aad:2015agg} & \cite{CMS-PAS-HIG-14-007} & \cite{CMS-PAS-HIG-14-013} & \cite{Aad:2015yza} & \cite{Aad:2014vgg} & \cite{Aad:2014fha} & \cite{Aad:2014cka} \\\hline
\multicolumn{1}{r|}{production $p = $}
& $gg$      & 2600 & 1200 & 320 & 38 & 38 & 32 & 11 & 7.7 & 5.1 & 1.5 \\
\multirow{8}{*}{$\displaystyle\left(\frac{{\rm BR}_i}{{\rm BR}_{\gamma\gamma}}\right)^{\rm max}$}
& $u\bar u$ & 1500 & 690 & 190 & 23 & 23 & 19 & 6.5 & 4.6 & 3.1 & 0.92 \\
& $d\bar d$ & 1600 & 700 & 200 & 23 & 23 & 20 & 6.7 & 4.7 & 3.1 & 0.94 \\
& $s\bar s$ & 2300 & 1000 & 280 & 34 & 34 & 28 & 9.6 & 6.8 & 4.5 & 1.4 \\
& $c\bar c$ & 2400 & 1100 & 300 & 36 & 36 & 30 & 10 & 7.3 & 4.8 & 1.5 \\
& $b\bar b$ & 2700 & 1200 & 340 & 40 & 40 & 34 & 11 & 8.1 & 5.4 & 1.6 \\
& $WW$      & 2400 & 1100 & 300 & 35 & 35 & 30 & 10 & 7.1 & 4.7 & 1.4 \\
& $ZZ$      & 2400 & 1100 & 310 & 37 & 37 & 31 & 10 & 7.3 & 4.9 & 1.5 \\
& $Z\gamma$ & 2200 & 980  & 270 & 33 & 33 & 27 & 9.2 & 6.5 & 4.3 & 1.3 \\
& $\gamma\gamma$ & 1700 & 760 & 210 & 25 & 25 & 21 & 7.1 & 5.0 & 3.4 & 1.0 \\
\hline
\end{tabular}
\end{center}
\caption{
Top: Bounds on 750~GeV resonances from 8~TeV LHC searches. Bottom: Derived bounds on ratios of branching fractions for different production channel assumptions. For $gg$ production, bounds on the branching fraction to $q\bar q$ are even tighter than indicated, since decays to $gg$ will necessarily also be present and enter the dijet searches. The same applies to branching fractions to $gg$ when the production is from quarks.\label{tab:othermodes}}
\end{table}

The ATLAS and CMS collaborations performed searches for resonant 
signals in many other final states as well. 
In table~\ref{tab:othermodes} we list the various two-body final states relevant 
to a neutral color-singlet spin-0 particle, and the corresponding $95\%$ C.L.\ exclusion limits, $(\sigma_8\times {\rm BR}_i)^{\rm max}$, from the 8~TeV searches. 
Searches for dijet resonances that employ $b$ tagging, and  would have enhanced 
sensitivity to $b\bar b$ final states, only address resonances heavier 
than $1$\,TeV~\cite{CMS:2012yf,CMS-PAS-EXO-12-023}, 
but the limits from $q\bar q$ searches still apply to $b\bar b$. 
The recent $13$\,TeV dijet searches~\cite{ATLAS:2015nsi,Khachatryan:2015dcf} 
do not cover the mass range around $750$\,GeV at all, due to triggering limitations.
We also note that the limits quoted in table~\ref{tab:othermodes} are approximate. 
In general, they do have some dependence on the width of the resonance, its spin, etc.

Table~\ref{tab:othermodes} also lists the resulting constraints on the 
ratios of branching fractions of the particle, for different production 
channel assumptions. 
They are computed as
\begin{equation}  \label{eq:brrat}
\left(\frac{{\rm BR}_i}{{\rm BR}_{\gamma\gamma}}\right)^{\rm max} = r_p\, \frac{(\sigma_8\times {\rm BR}_i)^{\rm max}}{\sigma_{13}\times\BRa}\,,
\end{equation}
where we use eq.~\eqref{sigmaBR13} and the cross section 
ratios $r_p$ from table~\ref{tab:ppaa}.

There is always a constraint from decays to dijets or dibosons since we take the resonance to 
couple to either $gg$, $q\bar q$, or the electroweak gauge bosons, for production. 
Also, the production cross section needs to be relatively large to accommodate 
the excess without too large $\ca$, so limits on dijet or diboson resonances may restrict part of the
parameter space of a concrete realisation.
For the case in which a single production channel dominates, we 
obtain upper limits on $\BRp$ and $c_p$ by saturating the corresponding dijet or diboson bounds:
\begin{equation}
\begin{split}
\BRp &<\sqrt{n_p\frac{M}{\Gamma} \frac{(\sigma_8\times {\rm BR}_p)^{\rm max}}{x_\sS^{8,p}}}\approx
\{
\overset{gg            }{\rule{0pt}{1.45em}\smash{25}},\, 
\overset{u\bar u       }{\rule{0pt}{1.5em }\smash{19}},\, 
\overset{d\bar d       }{\rule{0pt}{1.5em }\smash{25}},\, 
\overset{s\bar s       }{\rule{0pt}{1.5em }\smash{99}},\, 
\overset{c\bar c       }{\rule{0pt}{1.5em }\smash{120}},\, 
\overset{b\bar b       }{\rule{0pt}{1.5em }\smash{190}},\, 
\overset{WW            }{\rule{0pt}{1.5em }\smash{76}},\, 
\overset{ZZ            }{\rule{0pt}{1.5em }\smash{94}},\, 
\overset{Z\gamma       }{\rule{0pt}{1.45em}\smash{25}},\,
\overset{\gamma\gamma  }{\rule{0pt}{1.5em }\smash{4.2}}
\}\%
\times\left({\frac{45\,{\rm GeV}}{\Gamma}}\right)^{1/2} , \\
|c_p| &<\{
\overset{gg            }{\rule{0pt}{1.45em}\smash{97}},\, 
\overset{u\bar u       }{\rule{0pt}{1.5em }\smash{0.31}},\, 
\overset{d\bar d       }{\rule{0pt}{1.5em }\smash{0.35}},\, 
\overset{s\bar s       }{\rule{0pt}{1.5em }\smash{0.70}},\, 
\overset{c\bar c       }{\rule{0pt}{1.5em }\smash{0.79}},\, 
\overset{b\bar b       }{\rule{0pt}{1.5em }\smash{0.99}},\, 
\overset{WW            }{\rule{0pt}{1.5em }\smash{350}},\, 
\overset{ZZ            }{\rule{0pt}{1.5em }\smash{560}},\, 
\overset{Z\gamma       }{\rule{0pt}{1.45em}\smash{390}},\,
\overset{\gamma\gamma  }{\rule{0pt}{1.5em }\smash{110}}
\}
\times\left(\frac{\Gamma}{45\,{\rm GeV}}\right)^{1/4} .
\end{split}
\label{eq:cpupperbound}
\end{equation}
The dijet-excluded regions in the $c_p$--$\ca$ planes of figures~\ref{fig:cacg} 
and~\ref{fig:cacf} are the red-shaded areas.

By combining eq.~\eqref{eq:cpupperbound} with eqs.~\eqref{eq:BRprod} and \eqref{eq:cacp}, 
we obtain a second lower bound on
$\BRa$ and $c_\gamma$,
\begin{equation}
\begin{split}
\BRa &> 
\{
\overset{gg            }{\rule{0pt}{1.45em}\smash{0.98}},\, 
\overset{u\bar u       }{\rule{0pt}{1.5em }\smash{2.8}},\, 
\overset{d\bar d       }{\rule{0pt}{1.5em }\smash{3.5}},\, 
\overset{s\bar s       }{\rule{0pt}{1.5em }\smash{9.7}},\, 
\overset{c\bar c       }{\rule{0pt}{1.5em }\smash{11}},\, 
\overset{b\bar b       }{\rule{0pt}{1.5em }\smash{16}},\, 
\overset{WW            }{\rule{0pt}{1.5em }\smash{210}},\, 
\overset{ZZ            }{\rule{0pt}{1.5em }\smash{260}},\, 
\overset{Z\gamma       }{\rule{0pt}{1.45em}\smash{570}},\,
\overset{\gamma\gamma  }{\rule{0pt}{1.5em }\smash{400}}
\}\times 10^{-4}
\times 
\left(\frac{N}{20}\right)
\left(\frac{5.8~{\rm fb}^{-1}}{{\cal L}_{13}}\right) 
\left({\frac{45\,{\rm GeV}}{\Gamma}}\right)^{1/2} , \\
|c_\gamma| &> \{ {
\overset{gg            }{\rule{0pt}{1.45em}\smash{5.4}},\, 
\overset{u\bar u       }{\rule{0pt}{1.5em }\smash{9.2}},\, 
\overset{d\bar d       }{\rule{0pt}{1.5em }\smash{10}},\, 
\overset{s\bar s       }{\rule{0pt}{1.5em }\smash{17}},\, 
\overset{c\bar c       }{\rule{0pt}{1.5em }\smash{18}},\, 
\overset{b\bar b       }{\rule{0pt}{1.5em }\smash{22}},\, 
\overset{WW            }{\rule{0pt}{1.5em }\smash{80}},\, 
\overset{ZZ            }{\rule{0pt}{1.5em }\smash{88}},\, 
\overset{Z\gamma       }{\rule{0pt}{1.45em}\smash{130}},\, 
\overset{\gamma\gamma  }{\rule{0pt}{1.5em }\smash{110}}
} \}
\times \sqrt{\left(\frac{N}{20}\right)
\left(\frac{5.8~{\rm fb}^{-1}}{{\cal L}_{13}}\right) }
\left(\frac{\Gamma}{45\,{\rm GeV}}\right)^{1/4} .
\end{split}
\label{eq:cglowerbound}
\end{equation}
Depending on the width and the production mechanism, these bounds can be stronger or weaker than those in eqs.~\eqref{BRgmgmMin} and~\eqref{eq:calowerbound}.
Some comments are in order regarding the case of photon fusion
  dominance. In this case, eqs.~(\ref{eq:cpupperbound}) and (\ref{eq:cglowerbound})
  fix, for nominal width and signal strength, $|\ca| \approx 110$. This is because
  here we impose an upper bound of 2.5~fb for the 8~TeV diphoton signal, which
  essentially
  agrees with the predicted 8~TeV signal, for nominal width and number of excess
  events. For the same reason, this value agrees with the one previously
  obtained based on saturating the 13~TeV signal with a single diphoton coupling.

\begin{figure}[t]
\vspace{-1em}
\begin{center}
{\large $\boldsymbol{\Gamma=45}$ {\bf GeV}}\\
\hspace{-4mm}
\includegraphics[width=0.51\textwidth]{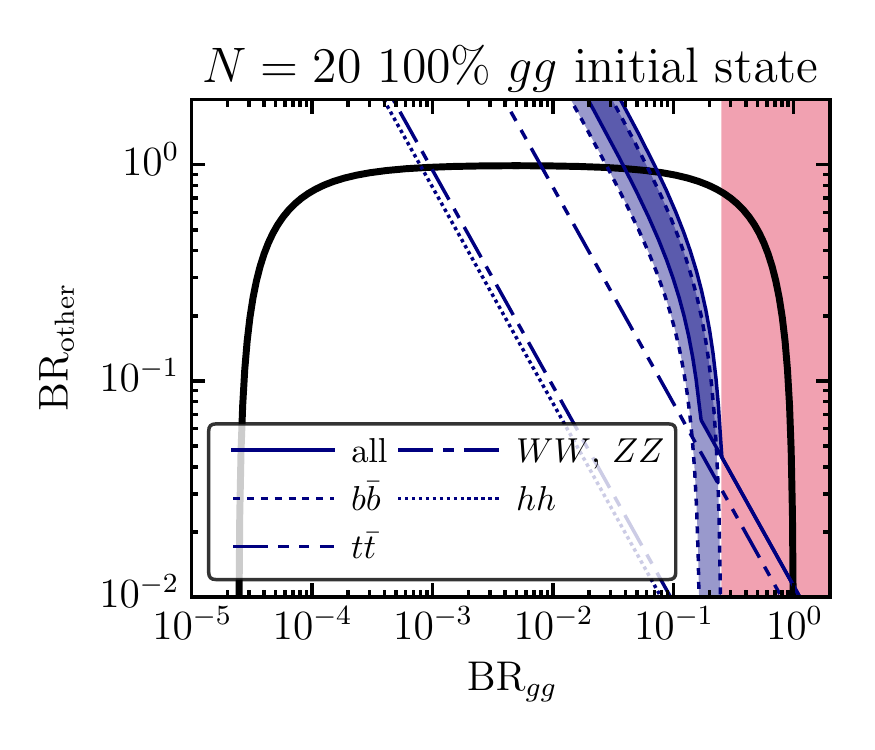}%
\includegraphics[width=0.51\textwidth]{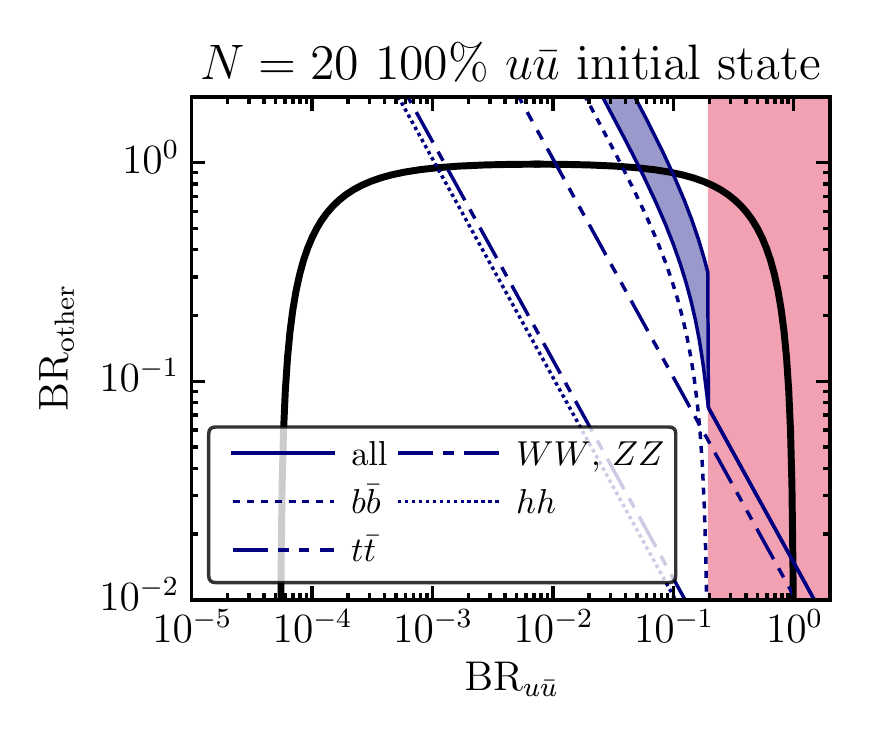}\\[-1.1em]
\hspace{-4mm}
\includegraphics[width=0.51\textwidth]{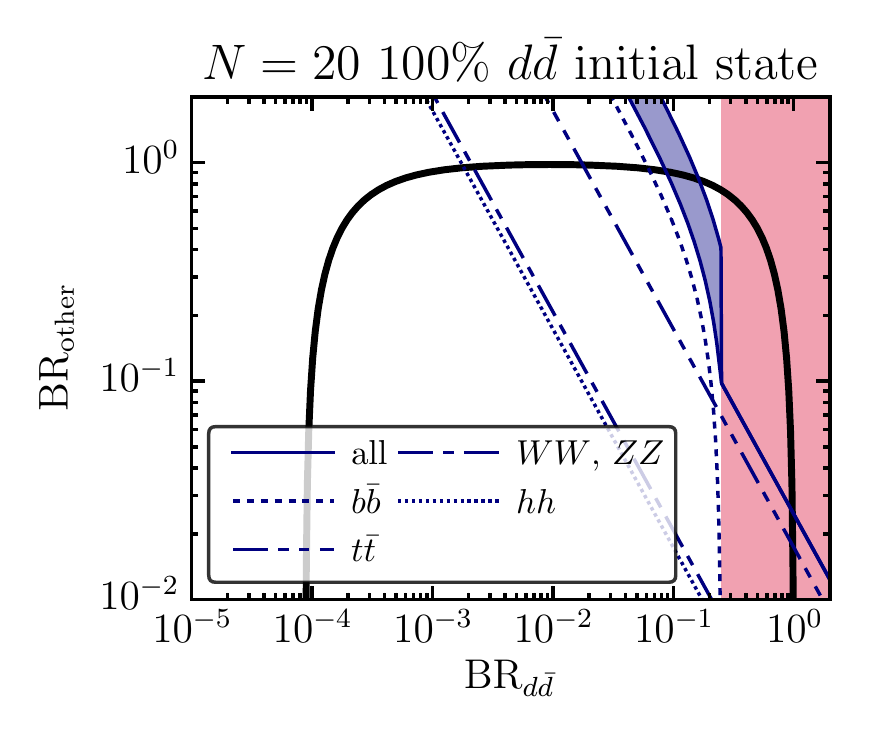}%
\includegraphics[width=0.51\textwidth]{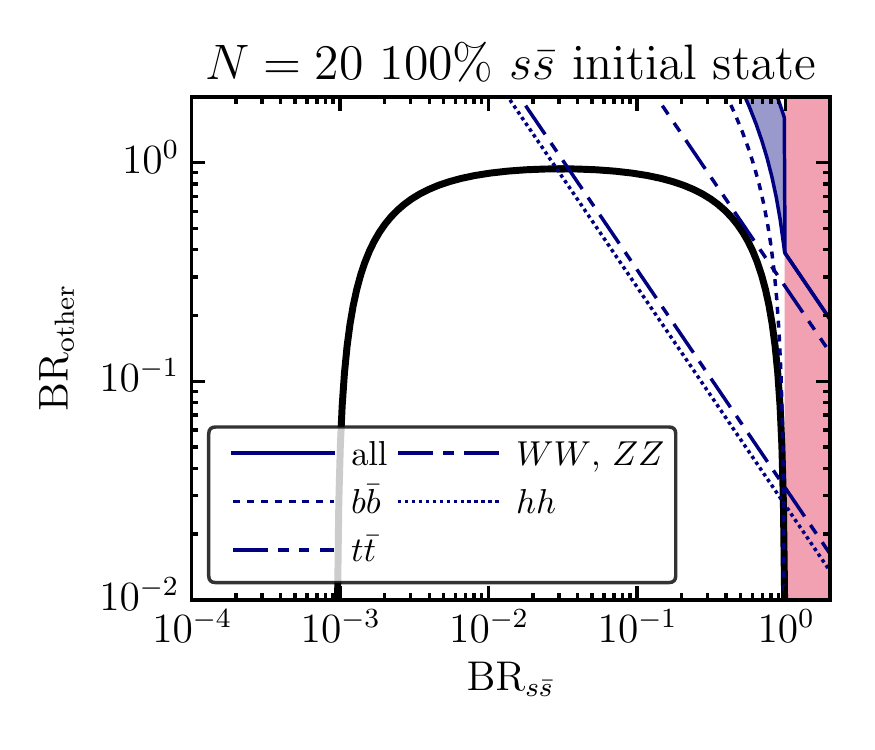}\\[-1.1em]
\hspace{-4mm}
\includegraphics[width=0.51\textwidth]{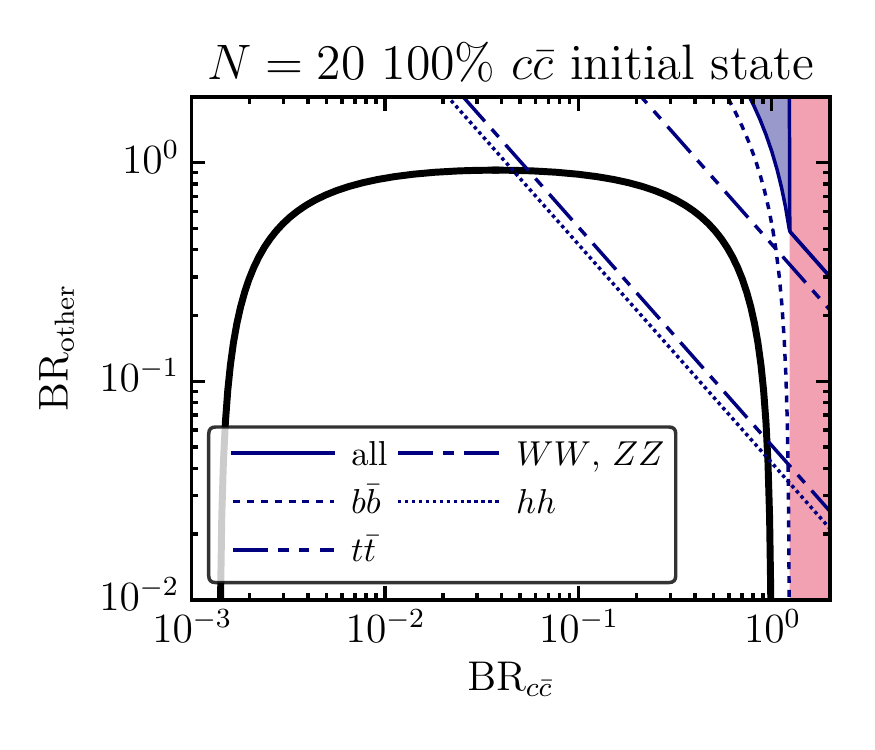}%
\includegraphics[width=0.51\textwidth]{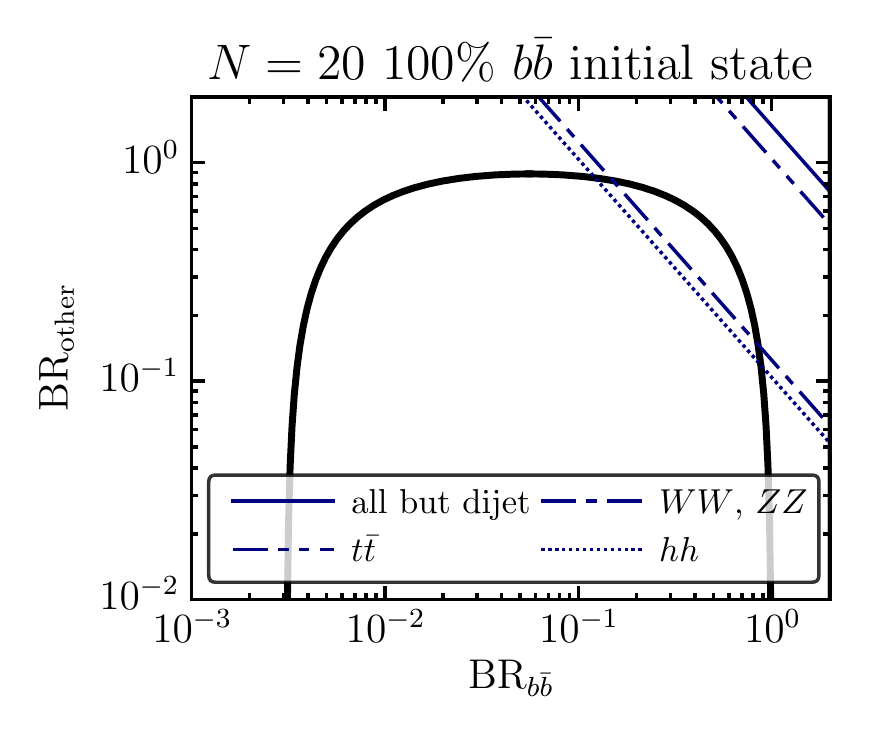}
\end{center}
\vspace{-2.8em}
\caption{The required branching fraction into modes other 
than the production mode and $\gamma\gamma$, $\BRu$, as a function 
of the production mode branching fraction, for $N = 20$ and $\Gamma = 45$\,GeV. 
Different plots correspond to different production mechanisms. 
Red regions are excluded by $8$\,TeV dijet resonance searches. 
Thin lines described in the legend show the maximal branching fractions allowed by $8$~TeV searches 
into final states from table~\ref{tab:othermodes}. 
The label ``all'' refers to the bound on the sum of all the final states 
from the table.
For mixed dijet final states ($gg$+$q\bar q$), we show bands extending between curves obtained using the $gg$ and the $q\bar q$ dijet constraint.
\label{fig:BRjjBRun-45}}
\vspace{-1em}
\end{figure}

\begin{figure}[t]
\vspace{-1em}
\begin{center}
{\large $\boldsymbol{\Gamma=1}$ {\bf GeV}}\\
\hspace{-4mm}
\includegraphics[width=0.51\textwidth]{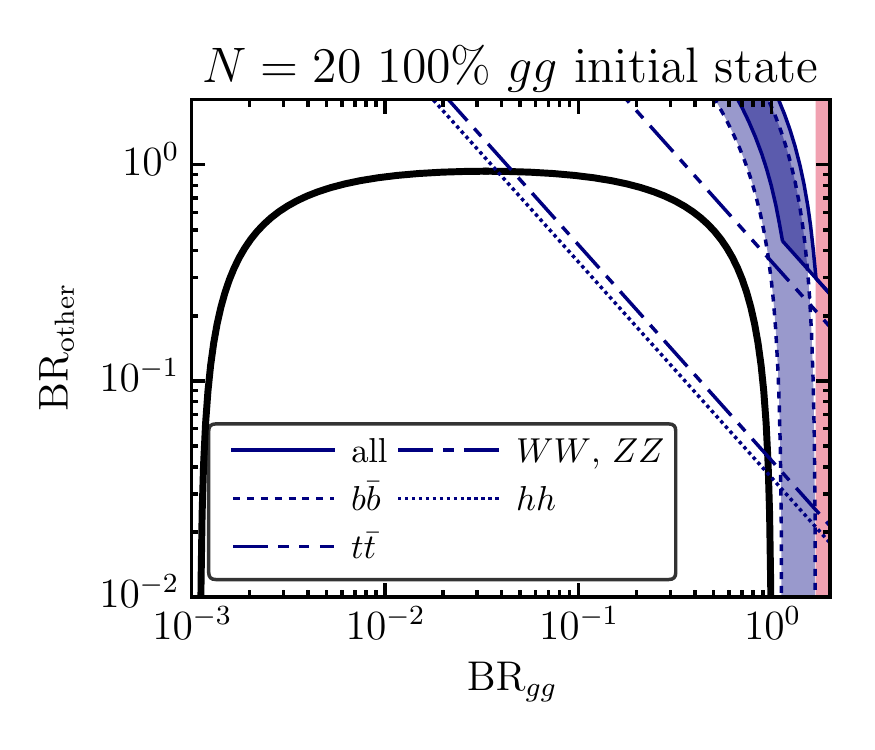}%
\includegraphics[width=0.51\textwidth]{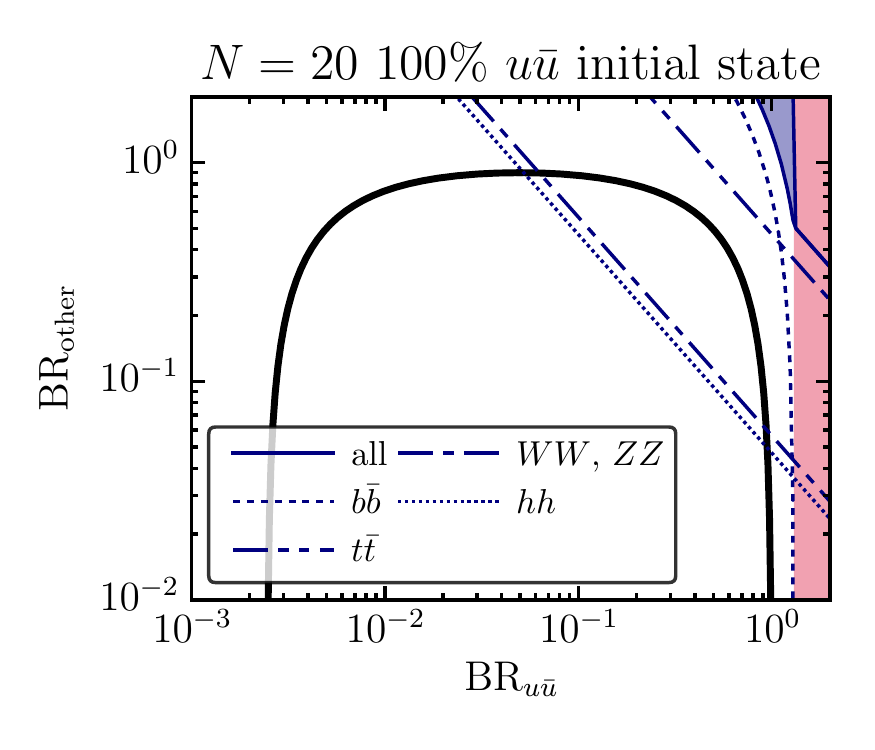}\\[-1.1em]
\hspace{-4mm}
\includegraphics[width=0.51\textwidth]{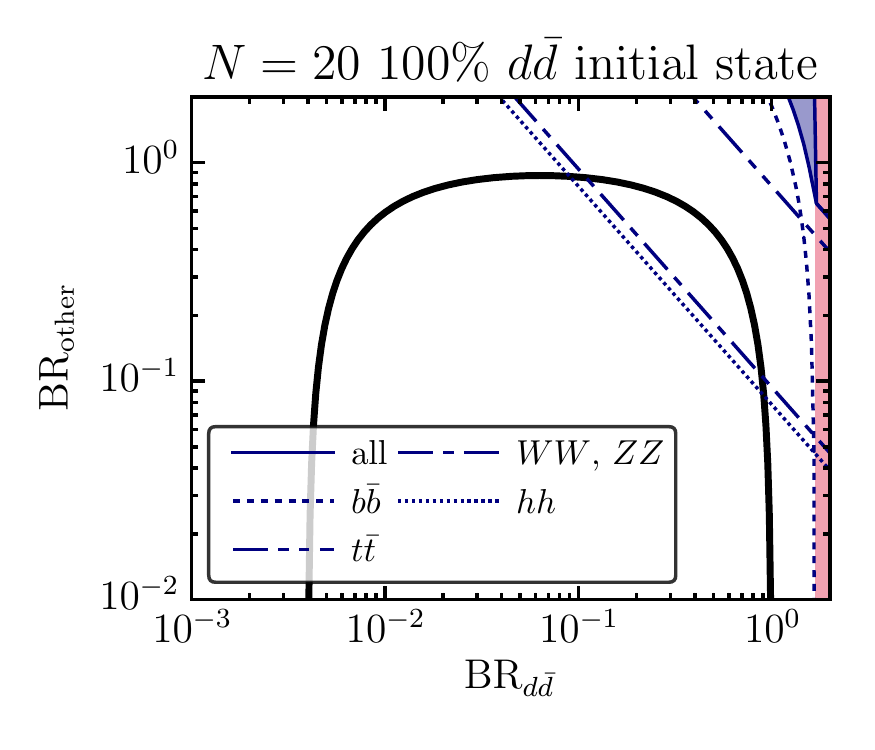}%
\includegraphics[width=0.51\textwidth]{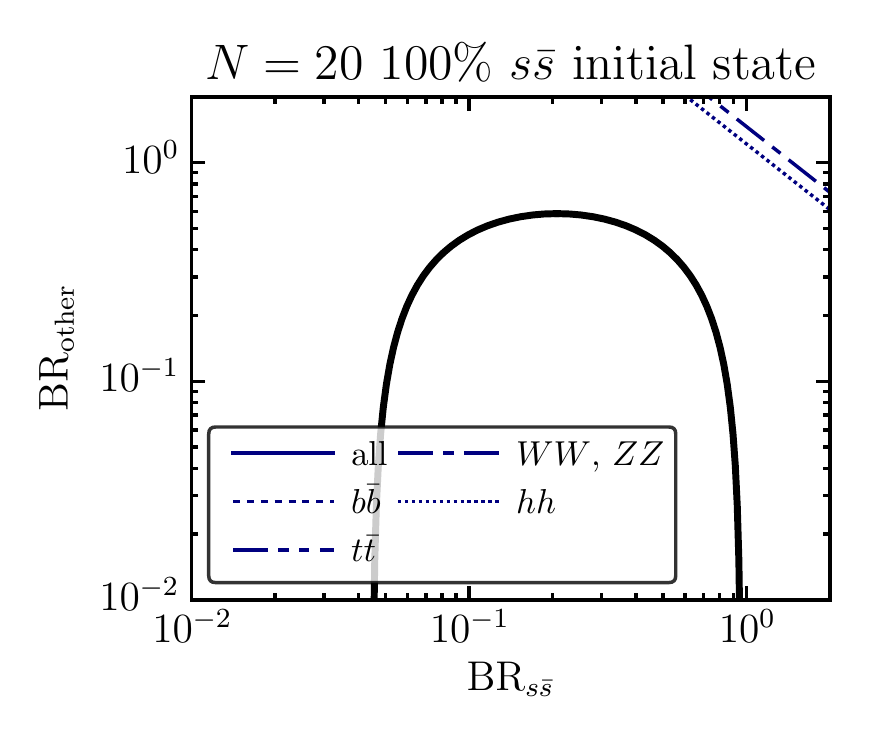}\\[-1.1em]
\hspace{-4mm}
\includegraphics[width=0.51\textwidth]{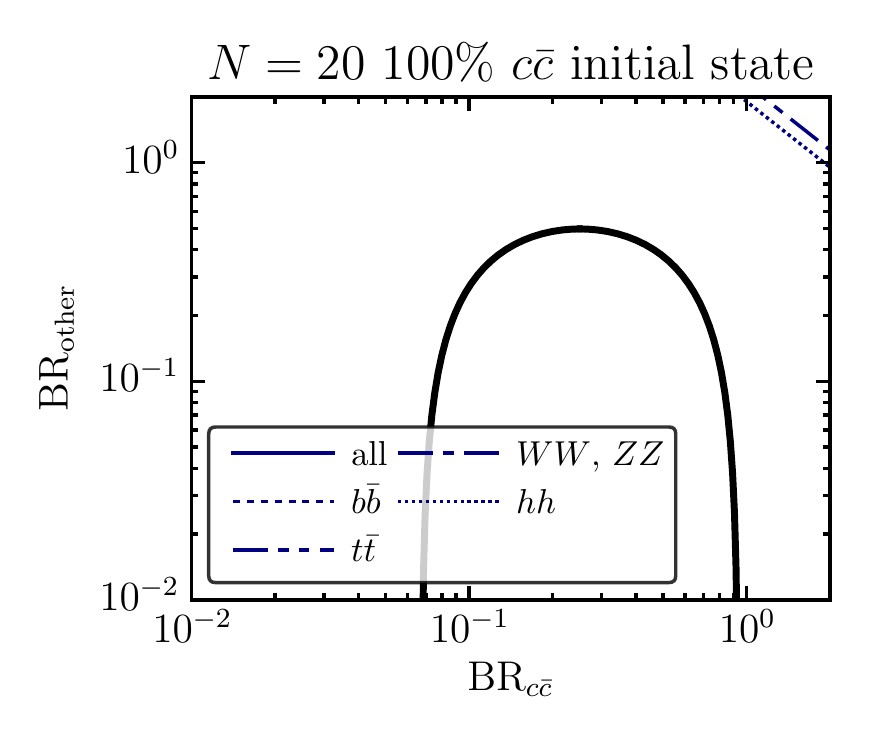}%
\includegraphics[width=0.51\textwidth]{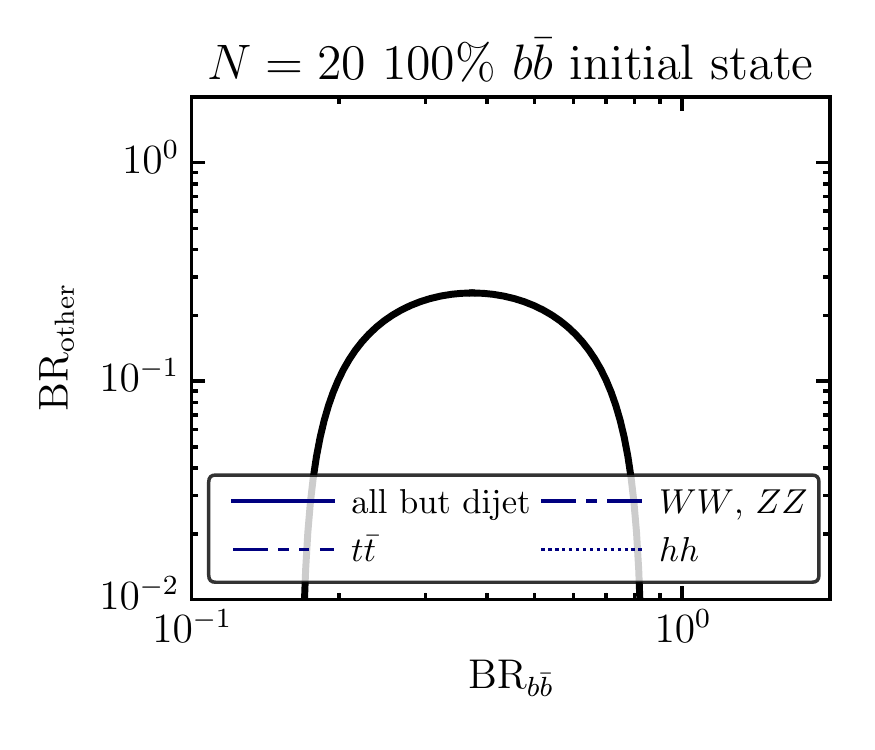}
\end{center}
\vspace{-2.8em}
\caption{The required branching fraction into modes other 
than the production mode and $\gamma\gamma$, $\BRu$, as a function 
of the production mode branching fraction, for $N = 20$ and $\Gamma = 1$\,GeV. 
Different plots correspond to different production mechanisms. 
Red regions are excluded by $8$\,TeV dijet resonance searches. 
Thin lines described in the legend show the maximal branching fractions allowed by $8$~TeV searches 
into final states from table~\ref{tab:othermodes}. 
The label ``all'' refers to the bound on the sum of all the final states 
from the table.
For mixed dijet final states ($gg$+$q\bar q$), we show bands extending between curves obtained using the $gg$ and the $q\bar q$ dijet constraint.
\label{fig:BRjjBRun-1}}
\vspace{-1em}
\end{figure}

Figures~\ref{fig:BRjjBRun-45} and~\ref{fig:BRjjBRun-1} show, for $\Gamma = 45$~GeV and $1$~GeV, respectively,
the required branching fraction $\BRu$ 
to modes other than the production mode and $\gamma\gamma$ as a function 
of the branching fraction of the production mode, $\BRp$. 
The black lines correspond to $N = 20$ signal events in the $13$\,TeV diphoton analyses.
These plots highlight the importance of $\BRu$, which in most of the viable
parameter space is the dominant branching fraction if the width $\Gamma$ is large. 
In blue lines, it is also shown to what extent $\BRu$ can be attributed 
to various decays into Standard Model particles, in view of the $8$\,TeV 
LHC bounds on such decays.
For example, if apart from the decays to the production mode and $\gamma\gamma$ the resonance
can decay only to $t \bar t$, the region above the corresponding blue line is
excluded. The solid blue
lines labeled ``all'' correspond to saturating all the two-body final states
listed in table~\ref{tab:othermodes}, with the band interpolating between lines that use the $gg$ and the $q\bar q$ dijet bounds.
The band is needed since the maximal possible ${\rm BR}_{\rm other}$ is generally achieved for a mixture of $gg$ and $q\bar q$ decays. 
Indeed, for a fixed ${\rm BR}_{gg}$, one can add decays to quarks.
For a fixed ${\rm BR}_{q\bar q}$ (for a given flavor $q$), one can add decays
to either gluons or other quark flavors,
but gluons are preferable since they are less constrained.
It is reasonable to expect the bound on such mixed final states to lie somewhere within the band.
The same discussion applies to the $b\bar b$ band in the fixed ${\rm BR}_{gg}$ case.

We see that when the diphoton signal is achieved by a large coupling to gluons/quarks 
and a small coupling to photons (right-hand side of the plots), 
it may be difficult to obtain $\Gamma = 45$\,GeV with decays to SM particles alone 
(if we neglect the possibility of large branching fractions to $\nu\bar\nu$ 
or multibody final states). 
On the other hand, in the case of a small coupling to gluons/quarks and a 
large coupling to photons (left side of the plots) there is no such limitation.

\section{Models}
\label{sec:models}
We now turn to discuss concrete models. 
First, in section~\ref{sec:singletscalar}, we discuss interpretations 
of the resonance as a scalar that is a singlet under the SM gauge group.
Next, in section~\ref{sec:2HDM}, we consider the possibility of an $SU(2)_L$ doublet.
Finally, in section~\ref{sec:mssm} we study whether the resonance can be a heavy Higgs of the MSSM.

\subsection{SM-singlet scalar}\label{sec:singletscalar}
The possible interactions of a real singlet scalar with the SM fields, 
up to dimension-five terms, are
\begin{align}
{\cal L}_{\rm singlet} &= (\mu \Phi+ \kappa_{H1} \Phi^2)H^\dagger H
\nonumber\\\nonumber
&+  \frac{\Phi}{f} \left(\kappa_g \frac{\alpha_s}{8\pi} G_{\mu\nu}G^{\mu\nu} + \kappa_Y \frac{\alpha_1}{8\pi} B_{\mu\nu}B^{\mu\nu} + \kappa_W \frac{\alpha_2}{8\pi} W_{\mu\nu}W^{\mu\nu} + \kappa_{H2} |D_\mu H|^2 + \kappa_{H3} |H|^4 \right) 
\\
&- \frac{\Phi}{f} \left(Y^d_{ij} H \overline{Q_i} d_j +  Y^u_{ij} \tilde{H} \overline{Q_i} u_j +  Y^e_{ij} H \overline{L_i} e_j+h.c.\right)\,.
\label{sing}
\end{align}
We first discuss the renormalizable scenario in which  only the terms
on the first line are present. 
Next we consider a, still renormalizable,  model
where the diphoton and digluon couplings $c_g$ and $c_\gamma$
 are generated by additional vectorlike fermions. We also analyze the
 pseudoscalar case, where the possible interactions differ in several
 important ways from \eq{sing}, as we will discuss.
We then turn to scenarios where the nonrenormalizable
  couplings on the second and third line are present, generated by physics
above the scale $M$ and resulting  in ``local'' contributions to the
couplings $c_g$ and $c_\gamma$.
We consider the dilaton scenario of ref.~\cite{Goldberger:2008zz}
(except that the dilaton is in addition to the Higgs)
in which the $\kappa_{g,Y,W}$ are related to the
$\beta$ functions in the low-energy effective theory.
Finally we discuss the possibility of production of the resonance by quarks due to 
the presence of the couplings in the last line of \eq{sing}.

\subsubsection{Renormalizable model}
We consider the case with only the renormalizable couplings in \eq{sing}. 
The $\mu$ term
induces mixing of $\Phi$ with the SM-like Higgs field, and
we obtain two mass eigenstates, ${\cal S}$ and the observed $125$\,GeV
Higgs $h$. 
This results in ${\cal S}$ having tree-level couplings
proportional to those of the SM Higgs but suppressed by a universal factor,
\bea
g^{{\cal S}}_i=s_{\alpha} g^{h_{\rm SM}}_i ,
\eea
where $s_\alpha\equiv \sin {\alpha}$, $\alpha$ being the mixing angle.
This mixing also modifies the couplings of the observed 125 GeV Higgs boson 
with respect to what they would be in the SM.
The modified couplings are scaled by $\cos \alpha$ with respect to their SM values.
We must thus have $s_\alpha \lesssim 0.2$ to ensure that these modifications
are compatible  with Higgs and particularly electroweak precision 
measurements~\cite{Robens:2015gla}. 
The coupling to the light quarks is thus negligible, therefore the production
must be gluonic. At one-loop level, SM particles  generate an
effective  $c_g$ and $c_\gamma$.
To get  the largest possible $c_g$ and  $c_\gamma$ we
take  $s_\alpha=0.2$ and obtain using the expressions for the top and $W$-loop contributions in ref.~\cite{Spira:1995rr}
\be
|c_g|=1.6\,, \quad |c_\gamma|=0.09 \,.
\ee
If we assume a $45$\,GeV width, we need $|c_g c_\gamma| \approx 530$ to accommodate the excess (cf.\ eq.~\eqref{eq:cacp}), so these numbers are far too small. Even if we allow for a smaller width, they still do not satisfy the bound $|c_\gamma| \gtrsim 2.7$ from \eq{eq:calowerbound}.

Clearly we need large contributions from BSM states to $c_\gamma$, and
either $c_g$ or the couplings $c_f$ to quarks, in eq.~(\ref{eq:lagrangian}), to explain the size of the excess.

\subsubsection{Boosting $c_\gamma$, $c_g$ with new vectorlike fermions}
 \label{vlikesec}

To investigate whether new colored and charged particles can generate  large enough  $c_{g,\gamma}$, we consider  
the minimal case of an additional vectorlike fermion, a triplet under QCD with electric charge $Q_f$, that couples to $\Phi$ as
\be
{\cal L}= -y_Q {\Phi} \bar{Q}Q-m_Q \bar{Q}Q .
\label{lagq}
\ee
The fermion loop generates $c_{g,\gamma}$. Any mixing of $\Phi$ with the Higgs doublet would dilute the vectorlike loop contributions (which would be generally larger than the SM loop contributions) to the diphoton and digluon couplings of the  mass eigenstate ${\cal S}$. Thus, we assume that the mixing, which is in any case constrained to be small, is negligible and the mass eigenstate is ${\cal S}=\Phi$. 

\begin{figure}[!t]
\centering
\includegraphics[width=0.47\textwidth]{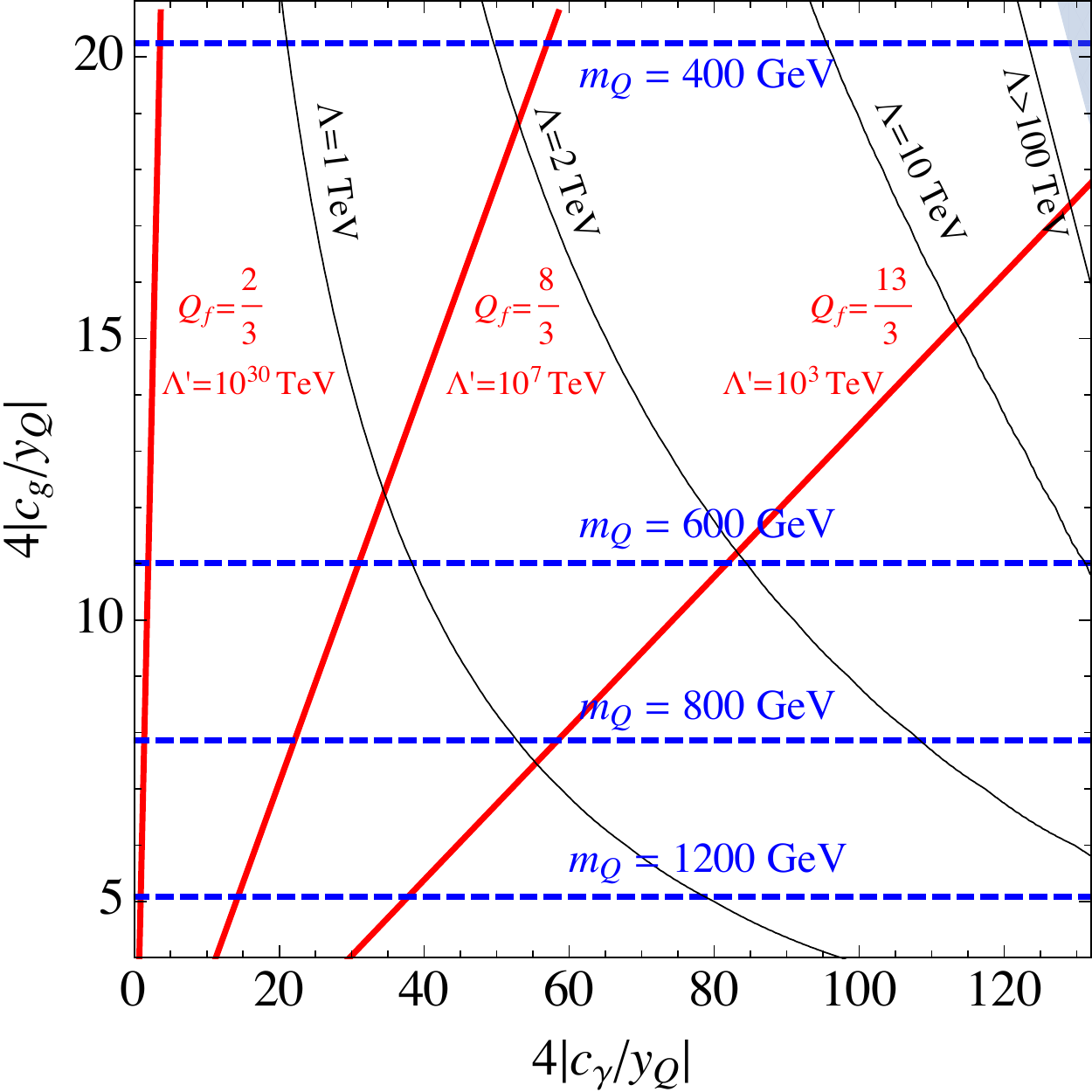}\quad
\includegraphics[width=0.47\textwidth]{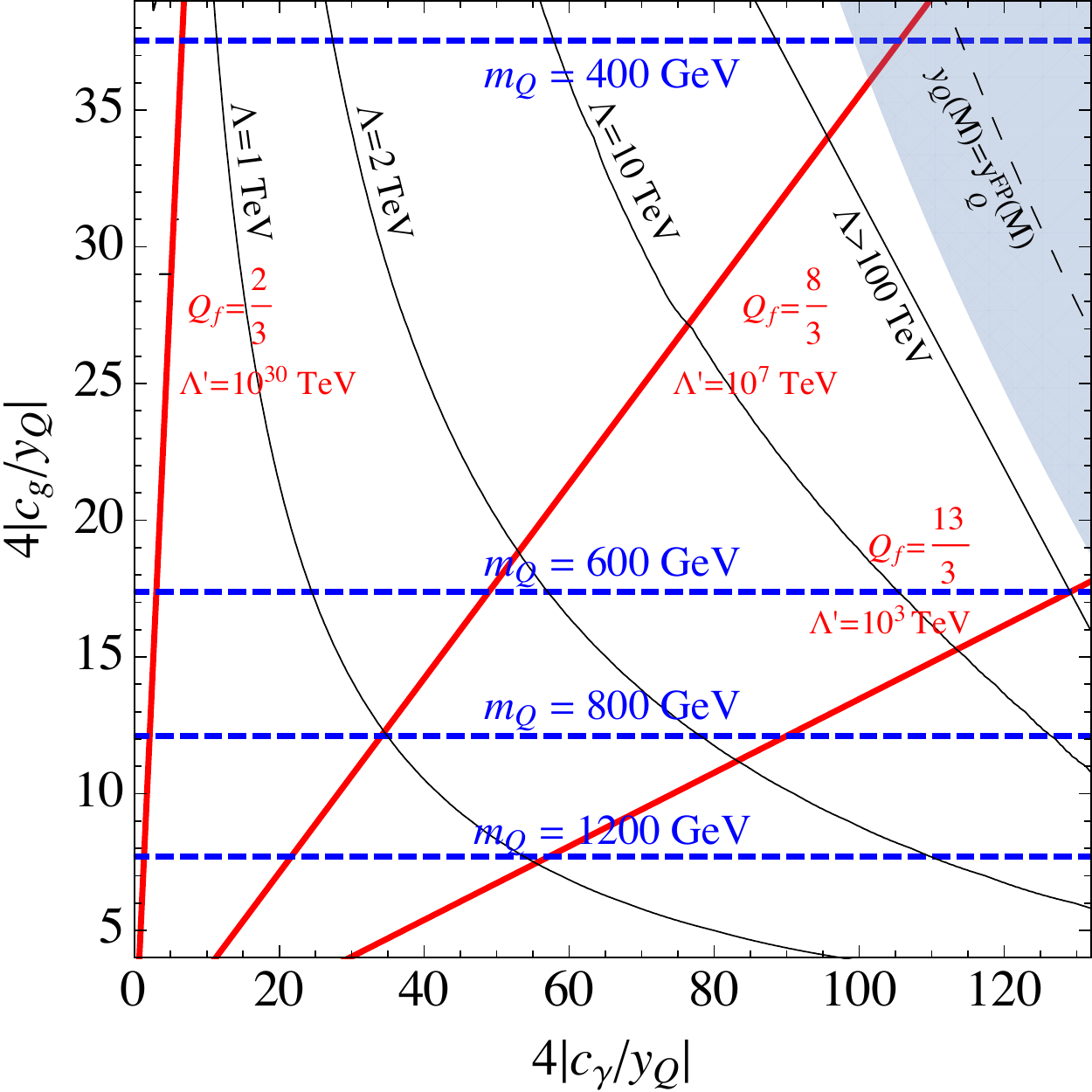}
\caption{
For a SM-singlet scalar (left) or pseudoscalar (right), contributions of a vectorlike color-triplet fermion with mass $m_Q$, charge $Q_f$ and Yukawa coupling $y_Q$ to the photonic, $c_\gamma$, and gluonic, $c_g$, couplings, scaled by $4/y_Q$.
Black lines are contours of the scale $\Lambda$ at which the theory becomes strongly coupled 
if the  value of $y_Q$   at the scale $M$ is fixed by requiring the correct signal size ($N = 20$), assuming $\Gamma = 45$~GeV.
The diagonal solid thick red lines stand for different values of $Q_f$; on each line, the corresponding scale for the Landau pole for the hypercharge interaction, $\Lambda'$ is shown. The horizontal blue dashed lines refer to different values of $m_Q$. We shade the regions where  $y^2_Q(M)$  is within 20\%  of  $[y_Q^{\rm FP}(M)]^2$ (see \eq{QFP}). The dashed black line is the contour where $y_Q(M)=y_Q^{\rm FP}(M)$.
\label{vectorlike}}
\end{figure}

The fermion $Q$ contributes~\cite{Spira:1995rr}
\begin{eqnarray}
   c_g &=& g_s^2 y_Q \tilde A_{1/2}(\tau_Q) , \\ 
   c_\gamma &=& 2 N_c Q_f^2 e^2 y_Q  \tilde A_{1/2}(\tau_Q) , 
   \label{cgca}
\end{eqnarray}
where $N_c = 3$ is the number of color states, $\tau_Q = M^2/(4 m_Q^2)$ and
\begin{equation}   \label{eq:Ahalftilde}
 \tilde  A_{1/2}(\tau)  = 2 \tau^{1/2} A_{1/2}(\tau)
                                       =  4 \tau^{-3/2} [\tau + (\tau-1) f(\tau) ] \,,
\end{equation}
where
\begin{eqnarray}
f(\tau)=\left\{
\begin{array}{ll}  \displaystyle
\arcsin^2\sqrt{\tau} & \tau\leq 1 \\
\displaystyle -\frac{1}{4}\left[ \ln\frac{1+\sqrt{1-\tau^{-1}}}
{1-\sqrt{1-\tau^{-1}}}-i\pi \right]^2 \hspace{0.5cm} & \tau>1 \,.
\end{array} \right.
\label{eq:ftau}
\end{eqnarray}
For $m_Q<M/2$ we obtain the constraint $y_Q \lesssim 0.7$ by requiring $\Gamma(\sS \to Q\bar Q) \lesssim 45$\,GeV. 
This would not allow generating sufficiently large values of $c_{g,\gamma}$. We thus take  $m_Q>M/2$. 
In figure~\ref{vectorlike} (left), we show the resulting $c_{g,\gamma}$ for a range of $m_Q$ and $Q_f$.  
The values of $c_{g,\gamma}$ for $y_Q=4$ can be directly read off from the plot, and one can easily find them for other 
$y_Q$ values by keeping in mind that $c_{g,\gamma}$ scale linearly with $y_Q$.

The same fermions will generically also generate couplings to $W^+W^-$, $ZZ$
and $Z\gamma$. While a detailed study of the various possibilities is beyond
the scope of this paper, we note that the bounds from
table~\ref{tab:othermodes} are easily satisfied if, for example, the
fermions are $SU(2)_L$ singlets. This is because they only contribute to the
$\kappa_Y$ coupling from eq.~\eqref{sing}, but not to $\kappa_W$, so one has
$\mbox{BR}_{Z\gamma}/\mbox{BR}_{\gamma\gamma} = 2\tan^2\theta_W \approx 0.6$,
$\mbox{BR}_{ZZ}/\mbox{BR}_{\gamma\gamma} = \tan^4\theta_W \approx 0.1$,
and no contribution to $W^+W^-$.

Since the Yukawa couplings $y_Q$ needed to reproduce the diphoton signal are
relatively large, it is important to check to what extent the theory remains
perturbative in the UV.
We first consider the case in which we assume a $45$\,GeV width for the resonance. 
In some regions of the parameter space, this implies a low cut-off for the theory at the scale at which $y_Q$ becomes strongly coupled. 
For $n_f$ color-triplet, $SU(2)_L$-singlet vector-like fermions, the RGE are given by\footnote{Note that the couplings  $\mu_3 \sS^3$ and $\lambda_S \sS^4$  do not alter these RG equations.  For a  more comprehensive analysis that considers the running of the quartic coupling $\lambda_S$, see for instance ref.~\cite{Son:2015vfl} where it is shown that  if the value of $\lambda_S$ at the low scale is appropriately chosen, it can remain perturbative up to  the Planck scale.}
{\allowdisplaybreaks
\begin{align}
\frac{d y_Q}{d \ln \mu} &= \frac{y_Q}{16 \pi^2} \left((3+6 n_f) y_Q^2- 6 Q_f^2 g'^{2}-8 g_s^2 \right)\,, \label{beyQ}\\
\frac{d g'}{d \ln \mu} &= \frac{g'^{3}}{16 \pi^2}\left(\frac{41}{6} +4 n_f Q_f^2\right)\,, \\
\frac{d g_s}{d \ln \mu} &= \frac{g_s^3}{16 \pi^2}\left(-7+\frac{2}{3} n_f\right)
\end{align}}
(see, e.g., ref.~\cite{Xiao:2014kba}); as said above we will only
consider the minimal case $n_f = 1$.
We show in figure~\ref{vectorlike} (left) contours of  the scale
$\Lambda$ at which the theory becomes strongly coupled, assuming $y_Q$
to be just large enough at each point in figure~\ref{vectorlike}
(left) to explain the excess. We take as the strong coupling scale
$\Lambda$ the scale where  either  $\sqrt{N_c}y_Q$  or (only
in some part of the region marked $\Lambda>100$ TeV) $\sqrt{N_c} Q_f g'$
becomes  ${\cal O}(4 \pi)$ ($N_c=3$).

For the theory to remain weakly coupled above the scale of roughly 10~TeV, a rather large value for the electric charge $Q_f$ is required, roughly above 3, for most of the shown parameter space.
For a large charge, the negative contribution to the running proportional to $y_Q Q_f^2 g'^{2}$ in eq.~\eqref{beyQ} can actually push the cut-off up to 100 TeV,
as shown in the top-right part of figure~\ref{vectorlike} (left). 
The RGE of $y_Q$ has a perturbative quasi fixed point, which in the one-loop approximation, eq.~\eqref{beyQ}, is given by
 \bea
 y_Q^{\rm FP}={1\over 3}\,\sqrt{8g_s^2 + 6 g'^2 Q_f^2} \approx 1.15 \sqrt{1 + 0.066 Q_f^2}\;.\label{QFP}
 \eea
Therefore, for an IR value (at the diphoton resonance mass scale) satisfying $y_Q<  y_Q^{\rm FP}$, the cutoff of the theory will likely be given by the Landau pole of the hypercharge interaction. It is controlled at high energies by the rather large charge of the vector-like quarks. 
On the other hand, for  $y_Q > y_Q^{\rm FP}$, the Yukawa coupling typically grows with energy. In such a case, 
the cutoff of the theory is set by the Landau pole of $y_Q$. 
We also note that generically, for UV boundary conditions that satisfy $y_Q(\Lambda_{\rm UV}) > y_Q^{\rm FP}(\Lambda_{\rm UV})$, we expect to have $y_Q^{\rm IR} \sim y_Q^{\rm FP}$. Since the one-loop $\beta$ function for $y_Q$ is small in this region, the impact of two-loop contributions may be nonnegligible. 
To indicate this, the region in which $y^2_Q(M)$ is within $\pm 20\%$ of $[y_Q^{\rm FP}(M)]^2$ is shaded.

We note that after rescaling $y_Q$ appropriately to explain the signal, the partial width to photons and gluons never exceeds 5~GeV in the region $Q_f>2/3$, so significant decays to other final states are needed to explain the 45~GeV width. The dijet constraint $|c_g| \lesssim 97$ (see \eq{eq:cpupperbound}) is satisfied in the above region
assuming that there are no dijet final states other than $gg$.

\begin{figure}[!t]
\centering
\includegraphics[width=0.49 \textwidth]{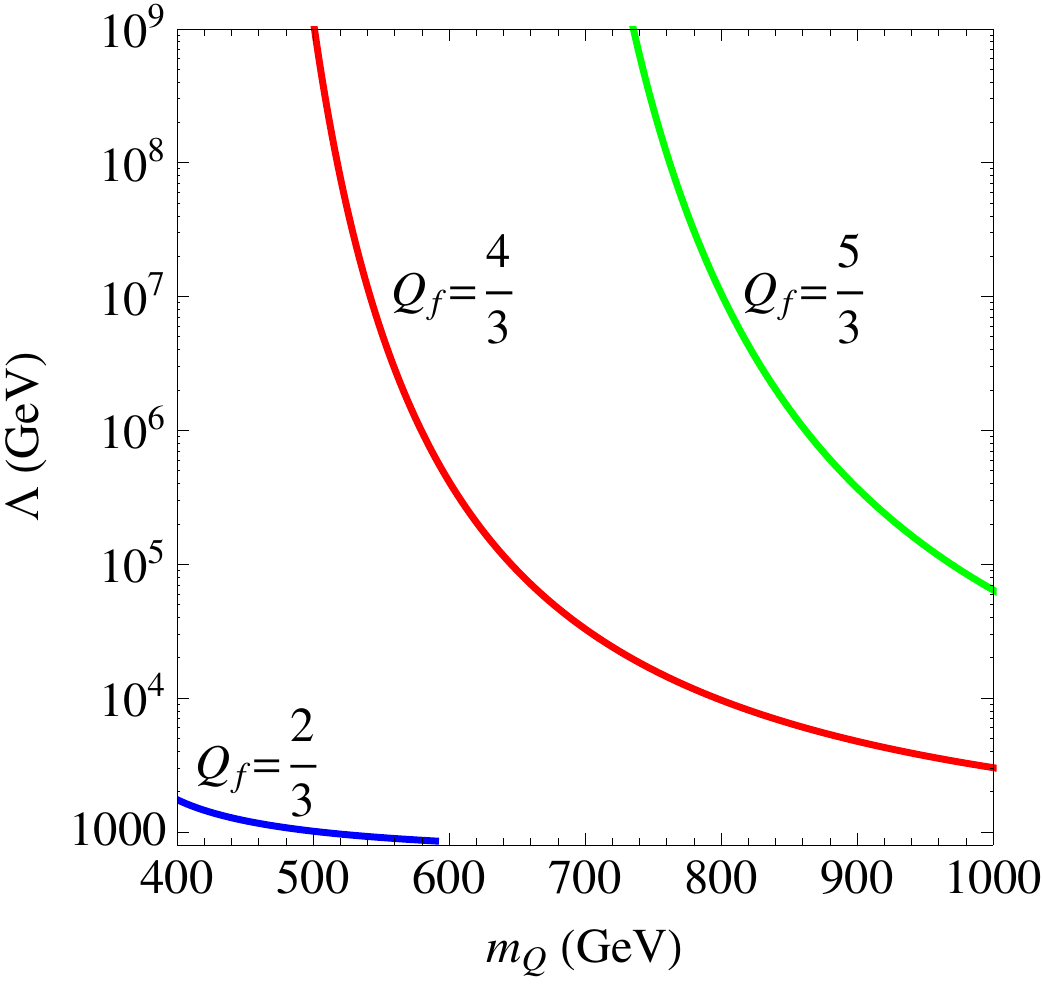}
\includegraphics[width=0.47 \textwidth]{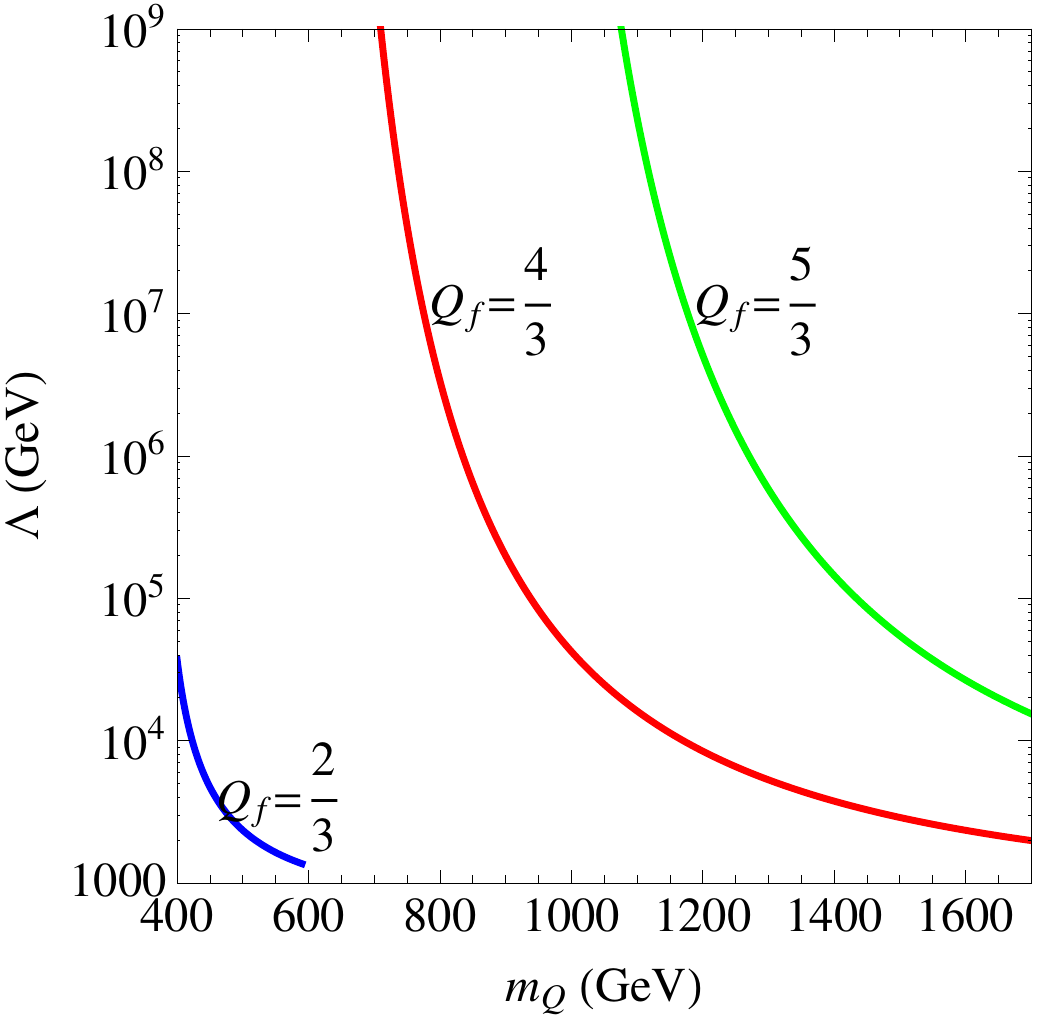}
\caption{
 For a narrow-width SM-singlet scalar (left) or pseudoscalar (right), the scale of breakdown of perturbativity as a function of the  color-triplet vectorlike fermion mass $m_Q$,
assuming we take the  required $y_Q$ at the scale $M$ to explain the excess.
For $m_Q\lesssim 490$~GeV ($m_Q\lesssim 720$~GeV)  with $Q_f =4/3$ ($Q_f =5/3$) for the scalar 
or $m_Q\lesssim 700$~GeV ($m_Q\lesssim 1050$~GeV) with $Q_f =4/3$ ($Q_f =5/3$) for the pseudoscalar, $y_Q (M) \lesssim y_Q^{\rm FP}$ in \eq{QFP} and the $\beta$ function is negative at the scale $M$, so the theory is perturbative.}
\label{nwavlike}
\end{figure}

We now consider the case in which the resonance is narrow and the width is dominated by decays to $gg$.
We then only need a $|c_\gamma| \approx 2.7$ according to \eq{eq:calowerbound}. 
In figure~\ref{nwavlike} (left)
we show the scale of breakdown of perturbativity assuming the required $y_Q$ to obtain 
$|c_\gamma| = 2.7$ from the loop of a vectorlike color-triplet fermion, as a function of its mass, $m_Q$.
We see that the theory might be perturbative up to
high scales even for  much smaller electric
charges.

\subsubsection{A pseudoscalar}

We now consider the case of a pseudoscalar.
Unlike the scalar it cannot mix with the SM Higgs and some couplings like those to longitudinal $W$'s and $Z$'s in \eq{sing} are not allowed. The possible interaction terms are
\begin{align}
{\cal L} &=  -\frac{1}{16\pi^2} \frac{1}{4} \frac{c_B}{M} \sS B^{\mu\nu}\tilde{B}_{\mu\nu} -\frac{1}{16\pi^2} \frac{1}{4} \frac{\cw}{M} \sS W^{a\mu\nu}\tilde{W}^a_{\mu\nu} -\frac{1}{16\pi^2} \frac{1}{4} \frac{\cg}{M} \sS G^{\mu\nu,a}\tilde{G}^a_{\mu\nu} \nn\\
&\quad -  \frac{\sS}{f} \left(iY^d_{ij} H \overline{Q_i} d_j +  iY^u_{ij} \tilde{H} \overline{Q_i} u_j + iY^e_{ij} H \overline{L_i} e_j+h.c.\right) -
y_Q  {\cal S} \bar{Q}i\gamma_5Q \nn\\
  &\supset
           -\frac{1}{16\pi^2} \frac{1}{4} \frac{\ca}{M} \sS F^{\mu\nu}\tilde{F}_{\mu\nu} 
           -\frac{1}{16\pi^2} \frac{1}{4} \frac{\cg}{M} \sS G^{\mu\nu,a}\tilde{G}^a_{\mu\nu} -
	   y_Q  {\cal S} \bar{Q}i\gamma_5Q \,,
\label{pseudo}
\end{align}
where we have also included a coupling  to a vectorlike quark $Q$.
A pseudoscalar can appear in composite models as a pseudo-Nambu-Goldstone boson (PNGB) with sizeable couplings to photons and gluons because of anomalies~\cite{Gripaios:2009pe}, but here we will consider the possibility where $c_g$ and $c_\gamma$ are generated only from loops of the fermion $Q$. These loop contributions are given by~\cite{Spira:1995rr}
\begin{eqnarray}
   c_g &=&2 g_s^2  y_Q \tilde A^{PS}_{1/2}(\tau_Q) , \\ 
   c_\gamma &=&  4 N_c Q_f^2 e^2 y_Q  \tilde A^{PS}_{1/2}(\tau_Q) , 
   \label{cgcaps}
\end{eqnarray}
where $\tau_Q = M^2/(4 m_Q^2)$ and
\begin{equation}   \label{aps}
 \tilde  A^{PS}_{1/2}(\tau)  = 2 \tau^{1/2} A^{PS}_{1/2}(\tau)
                                       =  2 \tau^{-1/2}  f(\tau) \,,
\end{equation}
with $f(\tau)$ defined in \eq{eq:ftau}.

Following the same procedure as in section~\ref{vlikesec},
we obtain the results shown in the right plots of figures~\ref{vectorlike} and~\ref{nwavlike}. They are qualitatively similar to the scalar case, but the theory can be perturbative up to somewhat higher scales for given $Q_f$ and $m_Q$.

Again, in figure~\ref{vectorlike} (right) we shade the region in which $y^2_Q(M)$ is within 20\% of $[{y_Q^{\rm FP}}(M)]^2$ of \eq{QFP}. This is the region where the cut-off can be high but at the same time our one-loop computation may be less reliable. The dashed black line is the contour where $y_Q(M)=y_Q^{\rm FP}(M)$.

We thus find that for a $\Gamma \approx 45$~GeV singlet resonance (in both the scalar and pseudoscalar cases) the size of the excess suggests
strongly coupled  physics at a few TeV unless there are additional new particles
around or below the mass of the  resonance 
with large electric charge. 
For the narrow width case we still need to require additional charged states but the theory can be perturbative with a smaller electric charge. 
The hints of strongly coupled physics motivate us to examine in more detail a popular strongly coupled scalar candidate, the dilaton.

\subsubsection{The dilaton}
\label{sec:dilaton}

We consider a generalization of the dilaton scenario of ref.~\cite{Goldberger:2008zz}, taking the full SM, including the Higgs doublet, to be part of a conformal sector (see also~\cite{Efrati:2014aea}). The dilaton is the PNGB of the spontaneously broken scale invariance.
The couplings of the dilaton in the electroweak broken phase 
are given by
\begin{align}
{\cal L}^{\rm dil}=&\frac{\Phi}{f} \Bigg( (\partial_\mu h_{\rm SM})^2 +2\left(m_W^2 W^+ W^-+\frac{m_Z^2}{2} Z^2\right)\nn\\
&\qquad -\sum_f m_f  \bar{f}f + \frac{\kappa_g \alpha_s}{8\pi} G_{\mu\nu}G^{\mu\nu}+\frac{\kappa_ \gamma \alpha}{8\pi} F_{\mu\nu}F^{\mu\nu} \Bigg),
\end{align}
where the first three terms arise from the operator $2 \Phi |D_\mu H|^2/f$. 
Note that the dilaton also couples to the $W^\pm$ and $Z$ field strengths, but these operators  have loop-suppressed Wilson coefficients and  thus their contribution is subdominant compared to the contribution from $2 \Phi |D_\mu H|^2/f$.  
Furthermore, there will be a mixing term with the SM Higgs, that will arise from the potential term $\Phi H^\dagger H$ and possibly also from kinetic mixing,  so that finally we obtain two mass eigenstates, ${\cal S}$ and the observed 125 GeV Higgs $h$, where
\be
{\cal S} = s_\alpha h_{\rm SM}+c_\alpha \Phi \,,
\ee
and  $s_\alpha={\cal O}(v/f)$ and $c_\alpha=1$ up to ${\cal O}(v^2/f^2)$.\footnote{Note that in the presence of kinetic mixing the transformation from $(\Phi, h_{\rm SM})$  to the  mass eigenstates is not orthogonal, and thus $s_\alpha$ and $c_\alpha$ cannot be expressed as a sine and cosine of a mixing angle (see for instance pp.\ 56-57 in ref.~\cite{Wells:2009kq}).}
We thus have for the couplings to the massive vector bosons and fermions
\be
g^{{\cal S}}_{V,f}=\xi g^{h_{\rm SM}}_{V,f} \,,\quad\text{with}~
\xi = s_\alpha +\frac{v}{f}c_\alpha \,.
\ee
For the dilaton, the couplings $\kappa_{g,\gamma}$ are completely determined using the low-energy theorems 
and scale invariance~\cite{Goldberger:2008zz}. The dilaton coupling to gluons is
\be
\frac{\Phi}{f}\,\frac{\sum_{\rm heavy} b_0^{i} \alpha_s}{8\pi} G_{\mu\nu}G^{\mu\nu} \,,
\ee
where  $b_0^i$ is the contribution of the field $i$ to the QCD $\beta$ function,
\be
\beta_i=\frac{b_0^i g_s^3}{16 \pi^2} \,,
\ee
and the sum is over all particles heavier than the scale $f$. Scale invariance implies
\be
\sum_{\rm heavy} b_0^{i}+\sum_{\rm light} b_0^{i}=0 \,,
\ee
so that we finally obtain
\be
\kappa_g = -\sum_{\rm light} b_0^{i} = 7 \,.
\ee
Similarly one obtains~\cite{Goldberger:2008zz}
\be
\kappa_ \gamma =-11/3 \,.
\ee
Note that if we do not include all the SM fields in the conformal sector 
but keep some of them elementary (e.g., ref.~\cite{Bellazzini:2012vz}), we cannot use the above arguments to fix $\kappa_ {\gamma,g}$ which then become model dependent.

The requirement $f \gtrsim M$ implies
\be
|c_g| \lesssim 21\,, ~~~|c_\gamma| \lesssim 0.7 \,,
\label{brs}
\ee
where we have assumed $s_\alpha \sim v/f$ in estimating the small contribution from mixing. For $f \approx M$, the total width, dominated by decays to $W^+W^-$, $ZZ$, $hh$ and $t\bar t$, is $\Gamma\approx 30$~GeV. For this width, \eq{eq:cacp} requires $|c_g c_\gamma| \approx 430$ to explain the excess. This cannot be obtained with the numbers in \eq{brs}. We also note that VBF production is negligible, considering the requirement in \eq{eq:calowerbound}.
  
Thus, we need additional large contributions to the QCD and QED $\beta$ functions below the scale $f$,
\bea
\Delta c_g &=&\frac{2 g_s^2 \Delta b_{\rm QCD} M}{f} \,,\\
\Delta c_\gamma &=&\frac{2 e^2 \Delta b_{\rm QED} M}{f} \,.
\eea
For $n_f$ additional vectorlike colour-triplet, $SU(2)_L$-singlet fermions  we have
\bea
\Delta b_{\rm QED}&=& \frac{4 N_c n_f}{3}Q_f^2 \,,\\
\Delta b_{\rm QCD}&=& \frac{2}{3}n_f \,,
\eea
where $Q_f$ is the electric charge of the fermion. Clearly to enhance $c_g$ and $c_\gamma$ to the extent that $|c_g c_\gamma| \approx 430$,
we need either  a very large charge $Q_f$ or a very large number of flavors $n_f$ of additional fermions below the TeV scale. This scenario thus appears contrived and we do not investigate it further.

\subsubsection{Production by  quarks}

Finally, we discuss the possibility of  production of ${\cal S}$ from quarks via the dimension-five
operators in \eq{sing}. Thus we consider the Lagrangian terms\footnote{Generating the coupling to photons via the $W^a_{\mu\nu}W^{a\mu\nu}$
operator instead of $B_{\mu\nu}B^{\mu\nu}$ would require a lower
cut-off.}
\be
{\cal L} \supset -\frac{{\cal S}}{f} \left(Y^d_{ij} H \overline{Q_i} d_j +  Y^u_{ij} \tilde{H} \overline{Q_i} u_j+h.c.\right) +  \kappa_Y \frac{\alpha_1}{8 \pi}\frac{{\cal S}}{f} B_{\mu\nu}B^{\mu\nu} \,.
\label{eft}
\ee
We want to find a conservative bound on the maximal energy scale up to which the EFT in \eq{eft} could be consistent while being completely agnostic about the UV theory.
We will consider scenarios in which ${\cal S}$ couples primarily to a
single quark flavor $f$  and set the corresponding $Y^{u,d}_{ij} \equiv Y_f$, as well as $\kappa_Y$, to their (conservative) perturbativity bounds as follows:
\be
\frac{Y_f}{f}\to \frac{16 \pi^2/\sqrt{N_c}}{\Lambda} \;,~~~~~~~~\kappa_Y \frac{\alpha_1}{8 \pi f} \to  \frac{4 \pi}{\Lambda}\;,
\label{qq-cutoff}
\ee
so that  $\Lambda$ can be
identified with the maximum scale up to which the theory could be predictive.%
\footnote{We can think of the following crude picture of how such a large diphoton coupling could be realised. Let us add a vector-like fermion $Q$ as considered in section~\ref{vlikesec}, but with mass $m_Q \sim \Lambda$, charge $Q_f \sim 4 \pi / g_1$ and Yukawa coupling to $\cal S$ of $y_Q \sim 4\pi$. This would generate the diphoton coupling in~\eqref{sing} with $f \to \Lambda$ and $\kappa_Y$ of order $(4\pi)^2/\alpha_1$, which indeed corresponds to the $4\pi$ in~\eqref{qq-cutoff}. This particular scenario does not require any exotic states below $\Lambda$, however generates a very large step in the hypercharge  $\beta$ function, which leads to a UV Landau pole for $g_1$, and hence for the entire Standard Model, at or close to the scale $\Lambda$.}
The couplings $c_f$ and $c_\gamma$  in \eq{eq:lagrangian} can be expressed in terms of $\Lambda$ as follows, 
\be
c_f= \frac{16 \pi^2}{\sqrt{N_c}}\frac{v}{\sqrt{2}\Lambda}\,,~~~~~~~~c_\gamma=256 \pi^3 \cos^2 \theta_W  \frac{ M}{\Lambda} \,.
\label{cfya}
\ee
One can find an absolute lower bound on $c_f$ by requiring that the production cross 
section of ${\cal S}$ is at least 6.9~fb in accordance with \eq{sigmaBR13}. 
For  an $f\bar f$ initial state  with $f=\{u,d,s,c,b\}$ this gives the bounds
\bea
\label{bnd2}
|c_u|\gtrsim 0.005 \;&\Rightarrow\; \Lambda \lesssim 3200~{\rm TeV}, \nn\\
|c_d|\gtrsim 0.007 \;&\Rightarrow\; \Lambda\lesssim 2300~{\rm TeV},\nn\\
|c_s|\gtrsim 0.022 \;&\Rightarrow\; \Lambda \lesssim 720~{\rm TeV}, \\
|c_c|\gtrsim 0.026 \;&\Rightarrow\; \Lambda\lesssim 610~{\rm TeV},\nn\\
|c_b|\gtrsim 0.040 \;&\Rightarrow\; \Lambda \lesssim 400~{\rm TeV}, \nn
\eea
respectively. From \eq{eq:calowerbound}, the coupling to photons, for an $f\bar f$ initial state, needs to satisfy
\bea
\label{bnd3}
|c_\gamma|\gtrsim &4.1&\;\Rightarrow\; \Lambda \lesssim 1100~{\rm TeV}, \nn\\
|c_\gamma|\gtrsim &5.2& \;\Rightarrow\; \Lambda\lesssim 880~{\rm TeV},\nn\\
|c_\gamma|\gtrsim &17&\;\Rightarrow\; \Lambda \lesssim 270~{\rm TeV},\\
|c_\gamma|\gtrsim &20& \;\Rightarrow\; \Lambda\lesssim 230~{\rm TeV},\nn\\
|c_\gamma|\gtrsim &30&\;\Rightarrow\; \Lambda \lesssim 150~{\rm TeV}, \nn
\eea
for $f=\{u,d,s,c,b\}$ respectively, thus  giving somewhat stronger bounds than \eq{bnd2}. 
The lower bound on $|c_\gamma|$ above can be saturated only in the
narrow width case with the additional requirement that $c_{f}$ is a
few times higher than the corresponding bound in \eq{bnd2} so that the
width is dominated by the decays to the production mode (see the
discussion below \eq{eq:calowerbound}). This would require that
$\Lambda$ is a few times lower than the bound in \eq{bnd2} which
roughly coincides with the values in \eq{bnd3}.

If we assume a 45~GeV width for the resonance, \eq{eq:cacp} must be satisfied, i.e.\ we must have
\bea
|c_\gamma c_u| &\approx& 2.9\;\Rightarrow\; \Lambda \approx 160~{\rm TeV}, \nn\\
|c_\gamma c_d| &\approx& 3.7\;\Rightarrow\; \Lambda \approx 140~{\rm TeV}, \nn\\
|c_\gamma c_s| &\approx&12\;\Rightarrow\; \Lambda \approx 80~{\rm TeV}, \\
|c_\gamma c_c| &\approx& 15\;\Rightarrow\; \Lambda \approx 70~{\rm TeV}, \nn\\
|c_\gamma c_b| &\approx& 22\;\Rightarrow\; \Lambda \approx 60~{\rm TeV}, \nn
\eea
for an $f\bar f$ initial state with $f=\{u,d,s,c,b\}$, respectively, where to obtain the values for the  cut-off we have used \eq{cfya}.

Note that, in the quark mass basis, the off-diagonal elements of the $Y$ 
matrices  generate terms like $c_{ij}{\cal S} \bar f_i f_j$ with $i \neq j$. 
Tree level FCNC constraints (see, e.g., ref.~\cite{Gupta:2009wn}) constrain these off-diagonal $c_{ij}$ to be  $\lesssim {\cal O} (10^{-4})$ 
for couplings involving the first two generations and $\lesssim {\cal O} (10^{-3})$ for couplings involving the $b$ quark, thus much smaller than the values of the diagonal couplings in eq.~\eqref{bnd2}.  
This scenario would thus be interesting from a flavor model-building point of view 
as one must find a way to suppress the off-diagonal couplings with respect to the 
diagonal ones.
For instance, notice that if $\cal S$ is a complex scalar, the coupling $c_{ij}{\cal S} \bar f_i f_j$ has an accidental flavour symmetry
that  forbids additional flavor violation, i.e.\ $\cal S$
can be formally viewed as a flavon field that carries an $i$-$j$ flavor charge
and thus cannot mediate $\Delta F=2$ flavor violation. In such a case, any flavor violation  induced by this coupling is proportional to powers of the $\ca$ coupling and/or the SM Yukawas that do not respect this accidental symmetry. 
The $\Delta F=2$ flavor violation induced by  the coupling $c_{ij}$ would  thus  be suppressed by  loop factors and/or SM Yukawas.

Let us now assume that a mechanism for alignment exists thus eliminating any  tree level FCNC. In this case flavor violation can arise only at higher loop order.
If the production is dominated either by up or down-type $\cal S$ couplings
we can assume that only one of  $Y^{u}$ or $Y^{d}$ is non-zero and that it is aligned to the quark mass basis.   
For instance let us consider the case where in the down mass basis, the production 
is dominated by a single coupling of $\cal S$, e.g.\ $Y_d={\rm diag}(y_d,0,0)$.
In such a case flavor violation has to involve the CKM matrix, $V_{\rm CKM}$.
Spurionically, the flavor violating bilinear coupling between two quark doublets 
is given by $V_{\rm CKM}^\dagger Y_d^\dagger Y_d V_{\rm CKM}$.
This spurion needs to be squared in order to generate the most dangerous $\Delta F=2$ 
contributions, in this case to $D-\bar D$ mixing processes.
As $Y_d$ is the coefficient of a dimension-five operator in the unbroken phase, 
each coupling is accompanied with an $\cal S$ field and thus the term $ Y_d^\dagger Y_d$
is generated only at one loop by integrating out $\cal
S$. 
This holds similarly for the CKM insertions, which can only arise from internal $W$ lines.
Consequently, the leading contribution to $\Delta F=2$ (involving quark doublets) 
would be suppressed by a three-loop factor and is thus negligibly small. 
There are possible two-loop contributions (mixing doublet and singlet quarks) that are, 
however, suppressed  by the light-quark masses and are thus even smaller. 
Finally, an even stronger (and phenomenologically not necessary) protection 
is obtained by assuming alignment and $U(2)$ universality in the form of 
$Y_d={\rm diag}(y_d,y_d,0)$ or  $Y_u={\rm diag}(y_u,y_u,0)$. In such a case the  contributions arise solely
via the mixing with the third generation.

%%%%%%%%%%%%%%%%%%%%%%%%%%%%%%%%%%%%%%%%%%%%%%%%%%%%%%%%%%%%%%%%%%%%%%
\subsection{Excluding the general pure  2HDM}
\label{sec:2HDM}
In this part we discuss the possibility of explaining the excess within the framework 
of the general two-Higgs-doublet model (2HDM), assuming no additional states beyond the additional doublet.
  
It is useful to describe the theory in the so-called Higgs basis~\cite{Gunion:2002zf}, 
where only one of the two doublets, which corresponds to the SM Higgs, acquires a VEV. 
The SM-like Higgs doublet, $H_a$, has a VEV equal to 246 GeV and a CP-even component with 
exactly SM-like couplings, whereas the other doublet, $H_b$, which contains the heavy CP-even and CP-odd states, as well as the charged states,
has a vanishing VEV.   The coupling
\be
-\lambda_{V}(H_b^\dagger H_a)  (H_a^\dagger H_a)+h.c.
\ee
causes a misalignment between the Higgs basis and the CP-even mass basis~\cite{Gupta:2012fy} that is of 
order $\lambda_{V} v^2/ M^2$.
If $\lambda_V \lesssim {\cal O}(1)$ we are in the so-called decoupling limit and  can  think 
of the ratio $\epsilon\equiv  \lambda_{V}v^2/ M^2$ as our formal expansion parameter 
(see ref.~\cite{Gunion:2002zf} and references therein for relevant discussions). 
The above interaction term leads to couplings of the heavy CP-even scalar, $H^0$, and 
the pseudoscalar, $A^0$, to the electroweak gauge bosons, $VV$.  
At the same time, it  causes deviations from SM values in the $h^0 VV$ couplings, 
$h^0$ being the lighter CP-even state. 
The value of $\lambda_{V}$ is thus constrained by electroweak precision measurements. 
Using the expressions in ref.~\cite{Branco:2011iw} we find the constraint
\be
|\lambda_V| \lesssim 3 \,,
\label{lamv}
\ee
which shows that we are in fact in the decoupling limit as $\epsilon \lesssim 0.3$\,.

One interesting consequence of the fact that $v^2/M^2 \ll 1$, is that
the mass splitting between the neutral CP-even state, $H^0$, and the odd one, $A^0$, which  is due to the coupling
\be
-\frac{\lambda_5}{2} (H_a^\dagger H_b)^2+h.c. \,,
\ee
is  generically small,
 \begin{equation}
\delta m = \left|m_{H^0}-m_{A^0}\right| \sim \frac{|\lambda_5| v^2}{2M} \sim 40\,{\rm GeV}
 \end{equation}
for $\lambda_5=1$. 
As $\delta m$ is compatible with the width of the excess, one may contemplate the possibility that 
the observed signal actually arises due to the presence of these two neighbouring states.

We will now show that  the general pure 2HDM cannot account for the observed excess. 
We note that in the Higgs basis the couplings of the heavy states to the light quarks can differ  
from those of the SM Higgs, as was exploited in ref.~\cite{Ghosh:2015gpa}; this is because $H_b$ acquires 
no VEV and thus its couplings to the SM fermions do not contribute to their masses.  
In particular, the couplings of $H_b$ to the light quarks
might be as large as allowed by the 
model-independent constraints in figure~\ref{fig:cacf}.
Thus, we consider production through either quark-antiquark or gluon fusion. 
In addition, as the signal might be accounted for by the presence of either $H^0$ or $A^0$, or both, 
we should consider the production and decay of each of these. 
We emphasise that to be conservative we do not require the width to be equal to $45\,$GeV as 
the excess could be explained by two narrower states separated in mass
by a few tens of GeVs, 
which would be consistent with the reported diphoton spectrum.
We denote by $N_H$ and $N_A$ the number of events from the production and decay of 
$H^0$ and $A^0$, respectively. 
In the CP limit we can assume no interference between these two production modes. 

\paragraph{Gluon-gluon production} Assuming that the masses of $A^0$ and $H^0$ are less than $45$\,GeV apart, 
both states would contribute to the excess. 
For the total width of the resonance to not exceed $45$\,GeV, eq.~\eqref{eq:width}, it is necessary that
\be
Y^2 \equiv \sum_f  \beta_f c_{f}^2 \lesssim 0.5\,,
\label{widthbound}
\ee
where $c_f$ is the coupling of $H^0$ and $A^0$ to the SM fermions,
\be
-c_f \overline{f}_L (H^0+i A^0) f_R+h.c.
\ee
and $\beta_f=(1- 4 m_f^2/M^2)^{1/2}$,
with $m_f$ being the fermion mass. Taking into account the steep decrease of the fermion loop
  functions $\tilde A_{1/2}$ and  $\tilde A^{PS}_{1/2}$, defined respectively in eqs.~\eqref{eq:Ahalftilde} and~\eqref{aps}, with
  decreasing quark mass, we find that for a fixed partial width to fermions (and thus fixed
$Y^2$), the fermionic loop contributions to $\ca$ and $\cg$ are
maximized for  $c_t/c_{f'} \gg 1$, where $c_t$ is the coupling to the top and $c_{f'}$ are couplings to fermions other than the top.

It is possible to bound the  contributions from $A^0$ because, unlike
for $H^0$, its couplings to the photons and gluons are only due to fermion loops. The total number of events from pseudoscalar decays can be expressed using \eq{eq:BRprod} as
\begin{equation}
N_A = 8.0 \times 10^5 \times \mbox{BR}(A^0\to gg) \times \mbox{BR}(A^0\to\gamma\gamma) \times \frac{\Gamma(A^0)}{45~\mbox{GeV}} \,.
\end{equation}
Using the inequalities 
\begin{equation}
\mbox{BR}(A^0\to gg) < \frac{\Gamma(A^0\to gg)}{\Gamma(A^0 \to f\bar f)}\,,\qquad
\mbox{BR}(A^0\to \gamma\gamma) < \frac{\Gamma(A^0\to \gamma\gamma)}{\Gamma(A^0 \to f\bar f)}
\end{equation}
along with the condition $\Gamma(A^0) \lesssim 45~\mbox{GeV}$ we then obtain
\begin{equation}
N_A \lesssim 8.0 \times 10^{5} \times \frac{\Gamma(A^0\to gg) \times\Gamma(A^0\to \gamma\gamma)}{\Gamma(A^0\to f\bar f)^2}\,.
\label{na}
\end{equation}
The partial widths are given by
\begin{align}
\Gamma(A^0\rightarrow gg) & =  \frac{\alpha_s^2 M}{32\pi^3}
\left| \sum_f c_f \tilde A^{PS}_{1/2}(\tau_f) \right|^2 \,,\\
\Gamma(A^0 \rightarrow\gamma\gamma)&=
\frac{\alpha^2 M}{64\pi^3}
\left| \sum_f c_f N_c Q_f^2  \tilde A^{PS}_{1/2}(\tau_f)
\right|^2\,,\label{agg}
\end{align}
where $\tau_f = M^2/(4 m_f^2)$ and $\tilde A^{PS}_{1/2}$ is defined in \eq{aps}.
Taking $c_t\gg c_{f'}$, as explained above, one can now evaluate the upper bound in~\eq{na},
\be
N_A \lesssim 0.02\,,
\ee
where we have used $\Gamma({A^0}\to t\bar t) = \frac{3}{8\pi}\sqrt{1-4m_t^2/M^2}M c_t^2 \approx 0.11\, M c_t^2$. We thus conclude that the pseudoscalar contributions are negligibly small in this case. 

We must then attribute all 20 signal events to $H^0$ decays,
\begin{equation}
20=N_H< 1.8 \times 10^{4}~{\rm GeV^{-1}} \times \frac{ \Gamma(H^0\to gg)}{\Gamma(H^0\to f\bar f) }\times \Gamma(H^0\to \gamma\gamma) \,,
\end{equation}
where we have used  \eq{eq:BRprod} and  $\Gamma(H^0\to f\bar f) < \Gamma(H^0)$. Now, as above, we take $c_t\gg c_{f'}$ in
\be
\Gamma(H^0\rightarrow gg)  =  \frac{\alpha_s^2 M}{128\pi^3}
\left| \sum_f c_f \tilde A_{1/2}(\tau_f) \right|^2 \,
\ee
to maximise the ratio $\Gamma(H^0\to gg)/\Gamma(H^0\to f\bar f)$, which becomes independent of $c_f\,.$ Using $\Gamma(H^0\to \gamma\gamma) =1.99 \times 10^{-7} M |c_\gamma|^2$ from table~\ref{tab:partialwidth} we then obtain the requirement
\be
|c_{\gamma}| \gtrsim 66 \,.
\label{cgambound}
\ee
As we will soon show, such large values of $|c_{\gamma}|$ are impossible to obtain in a pure 2HDM.

\paragraph{Quark-antiquark production}
As argued above, in general the heavy states can have sizeable couplings to the first two generations. 
Ignoring possible severe constraints from flavor physics we find that the weakest bound is from production 
due to $u\bar u$. Again we bound the pseudoscalar contribution first. 
Using   \eq{eq:BRprod} we have
\be
N_A = 8.1 \times 10^{3}~{\rm GeV}^{-1} \times{\Gamma(A^0\to u\bar u)\, \Gamma(A^0\to \gamma\gamma)\over \Gamma(A^0)} 
\,.
\ee
As the up-quark loop contributes negligibly to $\Gamma(A^0\to \gamma\gamma)$ compared with the top-quark loop, 
the above expression is proportional to $c_u^2 c_t^2/Y^2$ assuming that all the other fermionic couplings are zero.
Keeping the bound from \eq{widthbound} in mind, this is maximised for $c^2_u = 0.25$ and  $c_t^2  = 0.28$. 
For these values the pseudoscalar contribution yields less than one event. 
Thus $H^0$ must account for all 20 events. From~\eq{eq:calowerbound} we have the requirement
\be
|c_{\gamma}| \gtrsim 4.1\,.
\label{cgambound2}
\ee

Let us now discuss whether $|c_{\gamma}|$ as large as that required by eqs.~\eqref{cgambound} and~\eqref{cgambound2} can be obtained by loops of charged particles for the $H^0$ in the pure 2HDM.   
In addition to fermionic loops, the couplings of $H^0$ to photons receive contributions from loops of $W^\pm$ and $H^\pm$.
In the Higgs basis the two couplings that can parametrize these contributions are $\lambda_V$, defined in \eq{lamv}, and
\be
-\lambda_{H^+} (H_a^\dagger H_b)  (H_b^\dagger H_b)+h.c. \,,
\ee
where the term proportional to $\lambda_{H^+}$ ($\lambda_{V}$) results in a coupling of  the $H^0$ to the charged Higgs ($W$). 
We take the maximal value of $\lambda_V$ allowed by 
electroweak precision constraints, $\lambda_{V} \approx 3$, as already mentioned above.  There is no analogous restriction on the value of $\lambda_{H^+}$.
To check whether it is possible to satisfy the requirement in eq.~\eqref{cgambound}, or at least the one in eq.~\eqref{cgambound2}, we have added up the loop contributions 
from the top quark, the $W$ and the charged Higgs (see, e.g., ref.~\cite{Spira:1995rr}) allowing for maximal constructive interference. 
To maximise  $|c_\gamma|$, we take the charged-Higgs mass to be as small as $M/2$, which can, for instance, be obtained with a  large value of $\lambda_5$.
For ${\cal O}(1)$ values of $\lambda_{H^+}$, 
the contribution of the charged-Higgs loop is very small compared to the dominant contribution from the top loop as it is suppressed by $m_W^2/m^2_{H^+}$. We get, for $\lambda_{H^+}=1$, $|c_{\gamma}| \sim 1.8$. We find that to satisfy even the bound $|c_{\gamma}| \gtrsim 4.1$  in eq.~\eqref{cgambound2} requires very large values of $\lambda_{H^+}$, above $16\pi^2/3$. For such large values of $\lambda_{H^+}$, a naive estimate tells us that the loop contributions are a third of the tree-level ones, so perturbativity is questionable.  Such large values of  $\lambda_{H^+}$ and  $\lambda_{5}$ are also ruled out if we require their contribution
to the running of $\lambda_{V}$ between the scales $M$ and $m_Z$ to be smaller than the electroweak precision bound (which applies to $\lambda_{V}(m_Z)$), 
that is if we require $\Delta \lambda_{V} \lesssim 3$ (see ref.~\cite{Branco:2011iw} for the RGE).  This rules out both gluon and quark initiated production as the bounds in eqs.~\eqref{cgambound} and~\eqref{cgambound2}   are impossible to satisfy.

Thus, we have verified that the general 2HDM, without any additional states,  cannot account for the observed anomaly.

%%%%%%%%%%%%%%%%%%%%%%%%%%%%%%%%%%%%%%%%%%%%%%%%%%%%%%%%%%%%%%%%%%%%%%
\subsection{The fate of the MSSM}\label{sec:mssm}
We now turn to the Minimal Supersymmetric Standard Model (MSSM).
As in the 2HDM, which in its type-II form is contained in the MSSM as a subsector,
the only candidate particles for the resonance in the MSSM
are $H^0$ and $A^0$.%
\footnote{We consider the R-parity-conserving MSSM, otherwise in principle one
could consider sneutrino candidates, which can be similarly constrained.
A resonant $\gamma\gamma$ signal can also arise within the MSSM from
the annihilation of a squark-antisquark near-threshold QCD bound
state, most famously the stoponium~\cite{Drees:1993yr}. However, based
on expressions from~\cite{Kahawala:2011pc}, the stoponium has
$|c_\gamma| \simeq \sqrt{(2^{21}\pi^5/3^6)\bar\alpha_s^3\alpha^2} \approx 0.4$,
while eq.~\eqref{eq:calowerbound} requires $|c_\gamma| \gtrsim 2.7$
even for the most favorable (but also a quite generic for stoponium) scenario
where the width is dominated by decays to the production mode,
$\Gamma_{gg} \simeq (16/81)\bar\alpha_s^3\alpha_s^2 M \approx 0.0033$~GeV.
One might also consider the gluinonium, whose
binding is much stronger, though annihilation to $\gamma\gamma$ is
loop-suppressed~\cite{Kauth:2009ud,Kahawala:2011pc}. However, pair
production of $M/2 \approx 375$~GeV gluinos would have been almost certainly
noticed by now.}
The most plausible production mechanism is gluon fusion,
due to the smallness of the $H_d$ doublet's Yukawa couplings to light quarks
and the fact that we are deep in the
decoupling regime, $M_{H^0} \gg m_Z$. 

As we have seen above, the 2HDM fails by a large margin to
accommodate the data. However, in the MSSM there are extra contributions to
the $H^0gg$ couplings from sfermions and
to the $H^0\gamma\gamma$ couplings from sfermions and charginos, in
addition to those already present in the 2HDM. The $A^0gg$ and
$A^0\gamma\gamma$ vertices receive no sfermion contributions at
one loop as a consequence of CP symmetry, though
they do receive contributions from charginos.\footnote{As in the rest
of this work, we assume $CP$ conservation. Without this assumption,
the gluonic and photonic couplings of some
superposition of the two heavier mass eigenstates $H_2$ and $H_3$
will receive sparticle loop contributions, so apart from a division of
the diphoton signal between $H_2$ and $H_3$ resonant contributions,
we do not expect qualitative changes to our conclusions.}
Considering first $H^0$ as a candidate, dimensional
analysis gives, for the contribution of the two stops, for
$M_{\rm SUSY} = 1$ TeV,
\be
   c_g  \sim 2 g_s^2 \times \frac{v M_{H^0}}{M^2_{\rm SUSY}} \sim 0.5
\ee
and
\be
   c_\gamma \sim 2 N_c e^2 \times \frac{v M_{H^0}}{M^2_{\rm  SUSY}}  \sim 0.1 .
\ee
Even allowing for similar contributions from other sparticles,
this suggests that, generically, $|c_g c_\gamma| < 1$,
which is nearly three orders of magnitude below what is required according to
eq.~\eqref{eq:cacp}. However, we must also contemplate
that the true resonance width could be smaller than the ``nominal'' 45 GeV. The
decay width of $H^0$ is dominated by tree-level decays into top and
bottom quarks, and is essentially determined in the MSSM as a function
of $\tan\beta$, with a minimum of about 2 GeV at $\tan\beta \approx 6$.
 Hence, eq.~\eqref{eq:cacp} can be recast as
\begin{equation}   \label{eq:cacpmssm}
   \frac{|c_\gamma c_g|}{\sqrt{\Gamma(\tan\beta)/(45\,\mbox{GeV})} }=
   \rho_g \approx 530 .
\end{equation}

The question is how large the left-hand side may be. First, a small numerator
could be partly compensated for by a factor of up to five due to the denominator.
Second, an MSSM spectrum could also be quite non-degenerate, with
hierarchies like  $m_{\tilde t_1} \ll M_{H^0} , \mu \ll m_{\tilde t_2}$; this
is in fact favoured by the observed Higgs mass. In particular, large
$\mu$ and/or $A$-terms and a light stop can lead to a parametric
enhancement $\sim \{\mu, A_t \}/m_{\tilde t_1}$ relative to the naive
estimates above. Third,
there could also be important contributions from sbottoms and
staus, as well as charginos, which brings in a large subset of the
MSSM parameters. A conclusion about the fate of the MSSM requires
a quantitative treatment, but a brute-force parameter scan is not really feasible
and in any case beyond the scope of this work.
Instead, the purpose of the rest of this section is to obtain simple yet
conservative bounds on all one-loop contributions over the entire
MSSM parameter space. 

First, we will impose $1 \leq \tan\beta \leq 50$. 
The reason is that in the decoupling limit the $H^0 t\bar t$ and $H^0 b \bar b$ couplings are
$\sqrt{2} m_t/(v \tan\beta)$ and $\sqrt{2} m_b/(v \cot\beta)$,
respectively,\footnote{For this subsection, we use a convention $v
  \approx 174$ ${\rm GeV}$.} 
which, outside the stated $\tan\beta$ range, implies a decay
width that significantly exceeds the width allowed by observations, cf.\
section~\ref{sec:lag}.
(Independently, such large couplings would lead to a Landau pole in
$y_t$ or $y_b$, and/or
strong coupling at low scales. 
Our lower limit on $\tan\beta$
also has very strong support from the
observed Higgs mass of 125 GeV, which we will not separately impose.) 
The key assumption will be the
absence of charge- and colour-breaking minima of the scalar
potential. This could in principle be relaxed to only require
metastability over cosmological timescales; we leave this aside for
future work. 
As we will see, this assumption is
sufficient to exclude the MSSM if the resonance interpretation is confirmed.

\subsubsection{Constraints from vacuum stability}
An essential role in our argument is played by the upper bounds on the $\mu$ parameter and
the soft trilinear terms that follow from requiring the absence of
charge- and colour-breaking minima of the MSSM scalar potential. 
The derivation of these bounds is well known
\cite{Frere:1983ag,Derendinger:1983bz,Casas:1996de,Rattazzi:1995gk,Hisano:2010re,Altmannshofer:2012ks,Carena:2012mw,
Altmannshofer:2014qha} and involves suitable
directions of the MSSM scalar field space. 
We employ five such directions
\begin{equation}
  T_L = T_R = H_u^0, \quad
  B_L = B_R = H_d^0, \quad
  T_L = T_R = H_d^0, \quad
  B_L = B_R = H_u^0, \quad
  {\cal T}_L = {\cal T}_R = H_u^0,
\end{equation}
(with all other scalar fields held at zero), of which the first two
are $D$-flat. 
The five bounds derived from these directions can be formulated in terms
of the stop, sbottom and stau masses, as:
{\allowdisplaybreaks
\begin{align}
   |A_t| &\leq \sqrt{3} \sqrt{m_{\tilde t_1}^2 + m_{\tilde t_2}^2  - 2 m_t^2 +
                \frac{M_{H^0}^2}{2} \left(1+c_{2\beta} \right) - \frac{m_Z^2}{2} (1-c_{2\beta})(1+c_{2\beta})^2   } \;, \\
   |A_b| &\leq \sqrt{3} \sqrt{m_{\tilde b_1}^2 + m_{\tilde b_2}^2 - 2 m_b^2 +
                \frac{M_{H^0}^2}{2}\left(1 -c_{2\beta}\right)
		- \frac{m_Z^2}{2} (1-c_{2\beta})^2(1+c_{2\beta})} \;, \label{eq:Abbound} \\
   |\mu|  &\leq \sqrt{1 + \frac{m_Z^2}{m_t^2} \sin^2\beta }
   \nonumber\\
& \quad        \times \sqrt{m_{\tilde t_1}^2 + m_{\tilde t_2}^2 - 2 m_t^2
                   + \frac{M_{H^0}^2}{2}\left(1- c_{2\beta}\right)
		   - \frac{m_Z^2}{2}\left(1 + c_{2\beta} - c_{2\beta}^2 + c_{2\beta}^3\right)
	   } \; ,\\
   m_b |\mu| \tan\beta  & \leq
         m_t \sqrt{\frac{\tan^2\beta}{R^2} + \frac{m_Z^2}{m_t^2} \sin^2\beta  } \nonumber\\
& \quad       \times  \sqrt{m_{\tilde b_1}^2+ m_{\tilde b_2}^2 - 2 m_b^2
       + \frac{M_{H^0}^2}{2}  \left(1 +c_{2\beta}\right)
       - \frac{m_Z^2}{2}\left(1 - c_{2\beta} - c_{2\beta}^2 - c_{2\beta}^3\right)
       } \label{eq:mbmubound}\; ,\\
   m_\tau |\mu| \tan\beta  & \leq
         m_t \sqrt{\frac{\tan^2\beta}{R_\tau^2} + \frac{m_Z^2}{m_t^2} \sin^2\beta  } \nonumber\\
& \quad       \times  \sqrt{m_{\tilde \tau_1}^2+ m_{\tilde \tau_2}^2 - 2 m_\tau^2
       + \frac{M_{H^0}^2}{2}  \left(1 +c_{2\beta}\right)
       - \frac{m_Z^2}{2}\left(1 - c_{2\beta} - c_{2\beta}^2 - c_{2\beta}^3\right) 
       } \; ,
      \label{eq:mtaumubound}
\end{align}
where $R \equiv m_t/m_b \sim 50$, $R_\tau \equiv m_t/m_\tau\sim 100$, and $c_{2\beta}\equiv\cos(2\beta)$.
Eq.~\eqref{eq:Abbound} also has an analogue for $A_\tau$, obtained by substituting $b \to \tau$.}

In these expressions we have kept the
exact dependence on $\beta$, but neglected small terms of order
$m_Z^4/M_{H^0}^4$ (i.e.\ we have taken the decoupling limit).
Also, we have employed tree-level mass relations; this can easily be
undone (for example, $m_b |\mu| \tan\beta \to y_b |\mu|$ on the
left-hand side of eq.~\eqref{eq:mbmubound}, 
and $m_t \sqrt{\tan^2\beta/R^2 + m_Z^2/m_t^2 \sin^2\beta} \to y_t \sqrt{y_b^2 + (g^2+g'^{2})/2}$ 
on its right-hand side).

We can combine the bounds into bounding functions
$\Phi_t$, $\Phi_b$, and $\Phi_\tau$ of the sfermion masses and $\beta$ only. Firstly,
\begin{equation}
   \Phi_t = \left\{ \begin{array}{cc}
      \hspace*{-8cm} 0, & \hspace*{-9cm} m_{\tilde t_1}^2 + m_{\tilde t_2}^2 - 2 m_t^2 +
      \frac{M_{H^0}^2}{2}  (1+c_{2\beta}) 
      - \frac{m_Z^2}{2} (1-c_{2\beta})(1+c_{2\beta})^2 <0,\\[3mm] 
     \sqrt{3} \sqrt{m_{\tilde t_1}^2 + m_{\tilde t_2}^2 - 2m_t^2 
                  + \frac{M_{H^0}^2}{2}  (1-c_{2\beta}) 
		   - \frac{m_Z^2}{2}\left(1 + c_{2\beta} - c_{2\beta}^2 + c_{2\beta}^3\right)},
          &  \quad \mbox{otherwise} .
       \end{array} \right.
\end{equation}
If the condition for $\Phi_t=0$ is satisfied,
there is no way to satisfy the $A_t$ constraint; setting the bounding function
to zero in this case will serve to effectively discard those
unphysical points below. Otherwise, $\Phi_t$ simultaneously
bounds both $|A_t|$ and $|\mu|$.
A similar function that simultaneously bounds $|A_b|$ and $\frac{m_b}{m_t} |\mu|
\tan\beta$ is provided by
\begin{equation}
   \Phi_b = 
     \sqrt{3} \sqrt{m_{\tilde b_1}^2 + m_{\tilde b_2}^2 - 2 m_b^2 
                  + \frac{M_{H^0}^2}{2}  (1-c_{2\beta}) },
\end{equation}
and an identical function $\Phi_\tau$ follows from this by
substituting $b \to \tau$.

\subsubsection{Conservative bounds on sfermion contributions}
In the notation of ref.~\cite{Djouadi:2005gj}, the sfermion contributions
to $c_g$ are given by
\begin{eqnarray}
   \left|\sum_f c_g^{(\tilde f)}\right| &=&
     \frac{g\,M_{H^0}}{2\,M_W} g_s^2 \left|A^{H^0}_{{\rm SUSY},\tilde f} \right| .
\end{eqnarray}
If the contribution of a sfermion to $c_g$ is known, the
corresponding contribution to $c_\gamma$ is given by
\begin{eqnarray}
  c_\gamma^{(\tilde f)} &=& 2 (e^2/g_s^2) N_c^{(f)} Q_f^2 c_g^{(\tilde f)}.
\end{eqnarray}
Explicitly,
\begin{equation}
  A^{H^0}_{{\rm SUSY},\tilde f} 
     = 4 \sum_{\tilde f=\tilde t,\tilde b,\tilde \tau} \sum_{i=1,2}
     \frac{g^{H^0}_{\tilde f_i \tilde f_i}}{M_{H^0}^2} h(\tau^{\tilde f}_i)
\end{equation}
where $\tau^{\tilde f}_i = M_{H^0}^2/(4 m_{\tilde f_i}^2)$ and
\begin{equation}
  h(\tau) = \tau A^{H^0}_0(\tau) =
     \left\{ \begin{array}{cc}
          \frac{\arcsin^2(\sqrt{\tau})}{\tau} - 1 & \qquad \tau \leq 1 , \\
          - \frac{1}{4\tau} \left(
  \ln \frac{1 + \sqrt{1 - \frac{1}{\tau}}}{1 - \sqrt{1 -
      \frac{1}{\tau}}} - i \pi \right)^2 - 1 & \qquad \tau > 1 .
    \end{array} \right.
\
\end{equation}
Consider first the stops. In the decoupling limit, their couplings to $H^0$ are
\begin{equation}
  g^{H^0}_{\tilde t_i \tilde t_i} = -\cot\beta m_t^2 + x_i \sin(2\beta) m_Z^2 
                                      \pm m_t \frac{\sin(2\theta_{\tilde t})}{2} (\mu + A_t \cot\beta ) ,
\end{equation}
where $\theta_{\tilde t}$ is the stop mixing angle, and the
  coefficients $x_i$ depend
on $\theta_{\tilde t}$ and $\beta$ and are always less than one in magnitude. Using
that $h \to 0$ for $\tau \to 0, \infty$ and $|h| \leq h(1)
\approx 1.47$, one easily shows that the first two terms lead to
maximal contributions to   $A^{H^0}_{{\rm SUSY},\tilde f}$ that are
bounded (in magnitude) by $2.74 \cot\beta$ and $0.03$, respectively. (Similar terms
for the sbottom and stau cases will be negligible.) The third term in
the coupling leads to
\begin{equation}
   A^{H^0}_{{\rm SUSY}, \tilde t} = \frac{\sin(2\theta_{\tilde t})}{2} \frac{4\, m_t (A_t \cot\beta + \mu)}{M_{H^0}^2} 
     \times (h(\tau_1) - h(\tau_2) ).
\end{equation}
Employing now the bounding function $\Phi_t$ from the previous
subsection, it is not difficult to show that
\begin{eqnarray}   \label{eq:gHttbound}
   \left| m_t \frac{\sin(2\theta_{\tilde t})}{2} (A_t \cot\beta + \mu) \right| \!&\leq&
         m_t \min\left( \! \frac{1}{2}  \Phi_t (1 + \cot\beta) , 
             m_t \frac{\Phi_t^2}{m^2_{\tilde t_2} - m^2_{\tilde t_1}} 
          \right) \nonumber 
   \equiv \frac{M_{H^0}^2}{4} B(\tau_1, \tau_2; M_{H^0}) .
\\
\end{eqnarray}
The first argument of the min function follows from $|\sin| \leq 1$,
the second makes use of the explicit formula for the stop mixing angle.
We then have
\begin{eqnarray}
  |A^{H^0}_{{\rm SUSY},\tilde t}| \leq B(\tau_1, \tau_2) \left|h(\tau_1) -  h(\tau_2) \right| .
\end{eqnarray}
The right-hand side is bounded and the physical parameter space is the
compact region $0 \leq \tau_i \leq \tau_i^{\rm max}$,
where $\tau_i^{\rm max} = M_{H^0}^2/(4 m_{\tilde t}^{\rm min})$  depends
on the experimental lower bound on the lighter stop mass.
Straightforward numerical techniques establish that
\begin{equation}
   |A^{H^0}_{{\rm SUSY},\tilde t}| \leq 3.37,
\end{equation}
where the maximum is obtained at $\tan\beta=1$, when one stop is at threshold ($m = M_H/2$)
and the other is relatively light. Below, we numerically obtain and
use the bounds as a function of $\tan\beta$. 
(We allow stop masses as light as
100 GeV in the scan, to escape any doubts related to for instance compressed
spectra where light stops might have escaped detection at the LHC).
The extremal point is generally ruled out by the observed Higgs mass $m_h = 125$ GeV, and very
unlikely to be consistent with LHC searches, but we are being
conservative. 

Analogous steps lead to bounds on the sbottom and stau
contributions. In this case, terms proportional to $m_{b, \tau}^2$ and
$m_Z^2$ in the Higgs-sfermion couplings lead to completely negligible
effects. For the remainder, we require a bound
\begin{eqnarray}   \label{eq:gHbbbound}
   \left| m_b\frac{\sin(2\theta_{\tilde b})}{2} (\mu - A_b \tan\beta) \right| &\leq&
         m_t \min\left( \frac{1}{2} \Phi_b \left[\cot\beta + \frac{\tan\beta}{R} \right] \right. \; , \nonumber\\
    && \qquad \left.         m_t \frac{\Phi_b^2}{m^2_{\tilde b_2} - m^2_{\tilde b_1}}
                 \left[ \cot\beta + \frac{\tan\beta}{R} \right]
                 \left[1 + \frac{1}{R} \right]
         \right) \nonumber \\
   &\equiv& \frac{M_{H^0}^2}{4} B_{\tilde b}(\tau_1, \tau_2; M_{H^0}) \; .
\end{eqnarray}
The resulting bound is most effective in the intermediate $\tan\beta$
region, counteracting the small denominator of eq.~\eqref{eq:cacpmssm} in
that region. The bound on the stau contribution, as a function of
$\tan\beta$ and the slepton masses, is identical to the sbottom one,
except for a missing colour factor (overcompensated in the photonic
coupling by a ninefold larger squared electric charge).

\subsubsection{Contributions from other particles and verdict}
\begin{figure}
\centerline{
 \includegraphics[width=100mm]{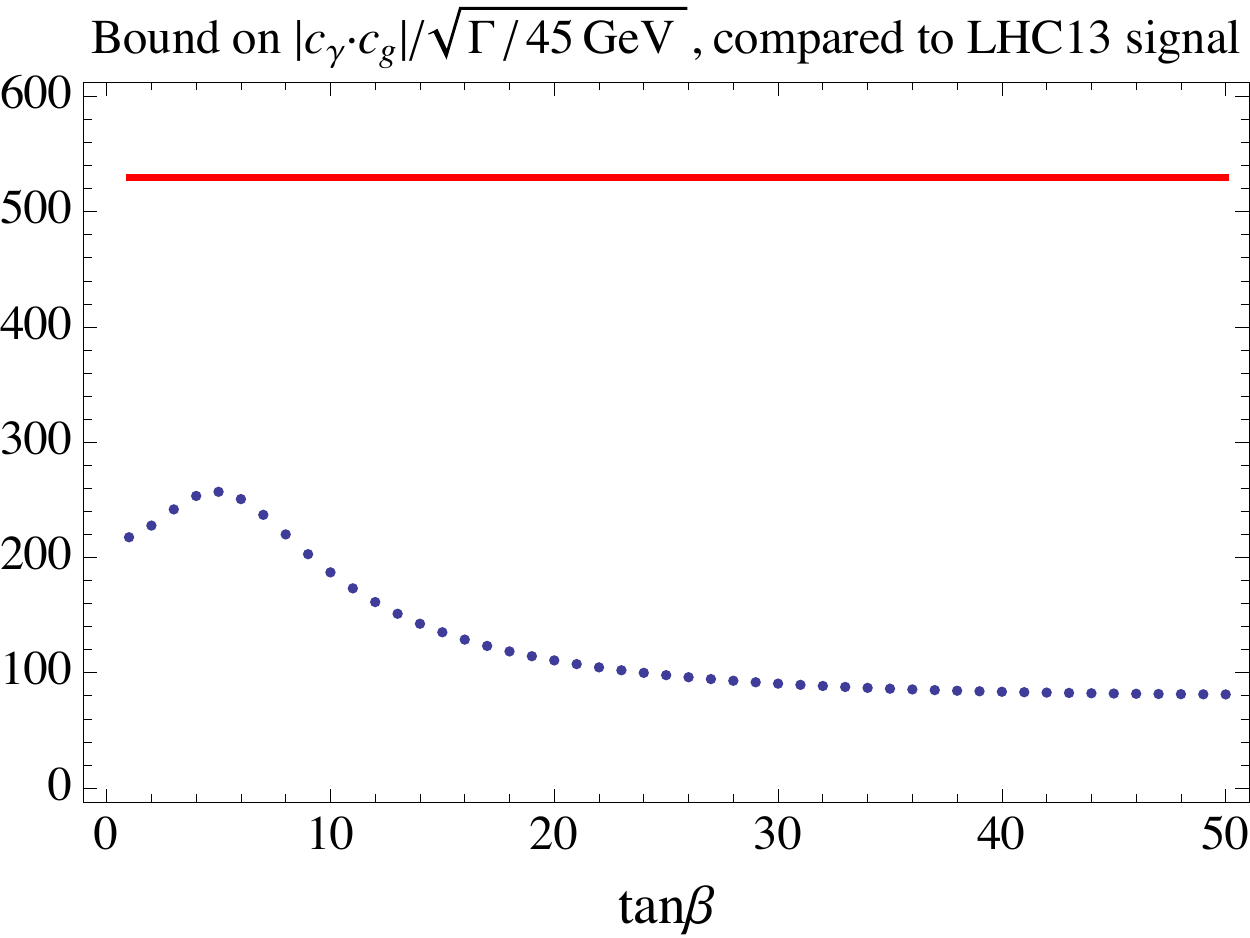} 
}
\caption{Comparison of the upper bound on the left-hand
    side of eq.~\eqref{eq:cacpmssm} to the
signal suggested by the diphoton excesses, as a function of
$\tan\beta$. The red horizontal line corresponds to
the signal, and the blue dots represent our conservative upper bound.
\label{fig:mssmvssignal}}
\end{figure}

The contributions from top,  bottom, $W$, and charged-Higgs loops
have already been discussed in the 2HDM section. In the decoupling
limit, where $M_{H^+} \approx M_{H^0}$,  they are essentially
functions of $\tan\beta$ only and easily incorporated.
Regarding charginos, their effect
is equivalent to the contribution of two vectorlike, colourless
particles; such contributions have also been discussed above. We only
need to bound the fermion loop function by its global maximum and make
no use of the relation of the chargino and Higgs mixing angles
to MSSM parameters in
order to obtain the bound $|c_\gamma^{\chi^+}| \leq 0.45$ (for any
$\tan\beta$). Assuming now the extreme scenario where all
contributions to $c_\gamma$ and $c_g$ simultaneously saturate their
bounds and are in phase with one another, we obtain a (very)
conservative upper bound on the left-hand side of
eq.~\eqref{eq:cacpmssm}. This is displayed in figure~\ref{fig:mssmvssignal}.
We observe that this bound still misses the data by more than a factor of two,
even at the point of closest approach at $\tan\beta \sim 5$. 
It is fairly clear that the bound could be made stronger by, for
example, employing more properties of the function $h$ or formulating
a higher-dimensional extremization problem (closer to a full scan of
the MSSM parameter space). 
It is also clear that the pseudoscalar $A^0$ fares
worse than $H^0$ as a resonance candidate: the chargino contribution to its coupling to photons is
similarly constrained as in the $H^0$ case, while
sfermion contributions to both the photonic and gluonic couplings are
absent, giving a much tighter bound on the left-hand side of eq.~\eqref{eq:cacpmssm}
in this case.

\subsubsection{Production from quarks?}{\label{sec:mssmqq}}
So far we only considered the production from gluons. 
A similar leading-order analysis for quark-antiquark initial states
again leads to a negative conclusion. 
The bounds just established translate to an upper bound
$| c_\gamma | < 5.3$, attained (for $\tan\beta \geq 1$) at $\tan\beta=1$. 
This can be combined with the model-independent analysis of section~\ref{sec:model-independent}.
First, the constraint in eq.~\eqref{eq:calowerbound} rules out initial states other
than $u \bar u$ or $d \bar d$. 
Eq.~\eqref{eq:cglowerbound} then implies $(\Gamma/(45 {\rm GeV}))^{1/4} < 0.5$, 
which together with eq.~\eqref{eq:cpupperbound} 
implies $|c_u| < 0.15$ for $u \bar u$ initial state 
($|c_d| < 0.18$ for $d \bar d$ initial state).
(The couplings $c_u$ and $c_d$ denote the
Yukawa couplings of the scalar mass eigenstate $H^0$, as defined
in section~\ref{sec:model-independent}. For finite $\tan\beta$
this state is a superposition of the neutral components of
$H_u$ and $H_d$, and yet another superposition of the doublets
in the ``Higgs basis'' of the preceding subsections.)
At the same time, the signal constraint 
together with the expression for the width-to-mass ratio
(eqs.~\eqref{eq:cacp} and \eqref{eq:width}, respectively)
imply 
\begin{equation}   \label{eq:cfboundmssm}
 |c_f| > \frac{2.9 (3.7)}{|c_\gamma|}
         \sqrt{(n_{\gamma\gamma} |c_\gamma|^2 + n_t |c_t|^2 + n_b |c_b|^2) \,\frac{750~{\rm GeV}}{45~{\rm GeV}} }\;.
\end{equation}
Using the tree-level relations $|c_t| = \frac{m_t}{\sqrt{2} v} \cot\beta$,
$|c_b| = \frac{m_b}{\sqrt{2} v} \tan\beta$, we find
this to be in conflict  with the upper bound unless 
$3 < \tan\beta < 15$ 
for $u \bar u$ initial state 
($4 < \tan\beta < 14$ 
for $d \bar d$ initial state),
in which case  $|c_u| > 0.10$ (or $|c_d| > 0.13$).
Employing again the tree-level relations,  these $\tan\beta$ ranges
correspond to an up-quark mass above 100 GeV (down-quark mass
above a few GeV), both in gross contradiction with observation.

However, higher-order corrections in the MSSM could potentially affect our conclusions. 
Although it is hard to see how they could give ${\cal O}(1)$ or larger corrections to the 
$H^0 g g$ or $H^0\gamma\gamma$ vertices, loop corrections can contribute  ${\cal O}(1)$
fractions of the down-type quark masses, through an induced coupling to the
doublet $H_u$.\footnote{%
We thank Martin Gorbahn for stressing this to us.} 
In this case, $c_b$ entering eq.~\eqref{eq:cfboundmssm} is no longer determined 
by $m_b$ and $\tan\beta$, and so for $\tan\beta \to \infty$ one would have only a very
weak bound $|c_u| > 0.005$ ($|c_d| > 0.007$) due to the partial width into diphotons.
While a complete investigation goes beyond the methodology and scope of this paper, 
we can put some relevant restrictions on such a scenario.

The fact that $\tau \tau$ resonance searches do not show an
excess results in an upper bound on the tree-level $\tau$
mass, giving
\begin{equation}
  m_\tau^{\rm tree} < \frac{7.4}{\tan\beta} {\rm GeV}.
\label{}
\end{equation}
This follows directly from the upper bound on the ratio
$\mbox{BR}_{\tau\tau}/\mbox{BR}_{\gamma\gamma} = (n_{\tau}/n_{\gamma})
(|c_\tau|^2/|c_\gamma|^2)$ (cf.\ eq.~\eqref{eq:brrat}  and
table~\ref{tab:othermodes}), using $|c_\gamma| > 5.3$, giving
$|c_\tau| < 0.026$. (This might be relaxed to about 0.03 for a mix of
$gg$ and $q \bar q$ production.)
This implies that either $\tan\beta < 10$ or the dominant fraction of
the $\tau$ mass would have to come from one-loop contributions.

Such one-loop contributions have been considered in the literature
(see, e.g., ref.~\cite{Borzumati:1999sp}) and are due to
neutralino-stau and chargino-sneutrino loops, with the latter
suppressed by the small $|y_\tau| = \sqrt{2} |c_\tau|/\sin\beta <
0.06$. Discarding them, the remaining neutralino-stau contributions
are proportional to the left-hand side of eq.~\eqref{eq:mtaumubound} 
times a combination of coupling constants, 
times a loop function.\footnote{%
If nonholomorphic soft terms are allowed, the left-hand side of
eq.~\eqref{eq:mtaumubound}
is modified but remains proportional to the relevant $\tilde \tau \tilde
\tau H_u^0$ coupling, such that the coupling remains bounded by the
right-hand side.}
For $\tan\beta > 8.3$, the $M_{H^0}$ and $m_Z$ dependence of the stability bound 
of eq.~\eqref{eq:mtaumubound} can be conservatively dropped. The
one-loop contribution of a given neutralino to the $\tau$ mass is then
bounded by the dimensionless combination
$$
  \sqrt{m_{\tilde \tau_1}^2 + m_{\tilde \tau_2}^2}\, m_{\chi_0}\,
    I(m_{\tilde \tau_1}^2, m_{\tilde \tau_2}^2, m_{\chi_0}^2)
$$
(with $I$ defined in~\cite{Borzumati:1999sp}), which is globally bounded in magnitude by one,
times a factor independent of sparticle masses. Summing the latter
over neutralinos and maximizing over mixing angles,
we find that $\Delta m_\tau^{\text{1-loop}} < 0.2$~GeV for $\tan\beta > 8.3$.
Therefore, if such a scenario can work at all, it necessarily
implies small $\tan\beta$. We leave a detailed investigation for
future work.

\subsubsection{Cautionary note}
We stress that our conclusions here are specific to the MSSM, and attest
to the high predictivity of the model. If the MSSM cannot survive in
regions of metastability (where charge and colour-breaking minima
exist but are not tunneled to over cosmological timescales),
or be saved by higher-order corrections,
more complicated supersymmetric models may still accommodate the excess,
although the techniques described here may be useful in scrutinizing
them. 
Another logical possibility of saving the MSSM would be
 production through the decay of heavier particles (say, stops,
 which could themselves be produced from gluino and squark decays).
As mentioned in the beginning, the experimental data do not seem to
support such a mechanism.

\section{Summary and Outlook}

This work deals with the core phenomenology of the diphoton excess observed by the LHC experiments
ATLAS and CMS around 750~GeV diphoton invariant mass. 
We have considered both the case where the data are interpreted by a narrow and a broad resonance.
We  obtained model-independent constraints on the allowed
couplings and branching fractions to various  final states, including the interplay with other existing bounds.
Our findings suggest that the anomaly cannot be accounted for by the presence of a single additional 
singlet or doublet spin-zero field and the Standard Model degrees of freedom; this includes all two-Higgs-doublet models.
We also found that, at least in a leading-order analysis, the
whole parameter space of the MSSM fails at explaining the excess 
if one requires the absence of charge and colour breaking minima.
If we assume that the resonance is broad,
we find that it is challenging to
find a weakly coupled explanation.
However, we provide an existence
proof in the form of a model with vectorlike quarks with
large electric charge.
For the  narrow resonance case, a similar model can be perturbative up to high scales also with smaller charges. 
We have also considered dilaton models where the full SM including the Higgs doublet is a part of the conformal sector.    We find that these models cannot explain the size of the excess unless
we add new fields  below the TeV scale to give large extra contributions to the QED and QCD beta functions.
As already mentioned, in all the scenarios studied by us we find that new particles below the TeV scale 
need to be present in addition to the resonance. 
They must have couplings to the scalar itself, to photons, maybe to gluons, and possibly also carry flavor information.
Further study of their LHC phenomenology would be interesting to follow.
Finally, models in which the new resonance has significant couplings to the light quarks motivate thinking about the linkage between flavor physics and the physics related to the resonance.

\vspace{-2mm}

\noindent
\\{\bf Note:} Other early-response studies of the various possible implications of the excess, 
that appeared approximately simultaneously with ours, are refs.~\cite{Harigaya:2015ezk,Mambrini:2015wyu,Backovic:2015fnp,Angelescu:2015uiz,Nakai:2015ptz,Knapen:2015dap,Buttazzo:2015txu,Pilaftsis:2015ycr,Franceschini:2015kwy,DiChiara:2015vdm,Higaki:2015jag,McDermott:2015sck,Ellis:2015oso,Low:2015qep,Bellazzini:2015nxw,Petersson:2015mkr,Molinaro:2015cwg}. Also, after the submission, an earlier study of diphoton resonances~\cite{Jaeckel:2012yz} was pointed out to us.

\vspace{-2mm}

\acknowledgments
GP is supported by the BSF, ISF, and ERC-2013-CoG grant (TOPCHARM \#
614794). SJ thanks GP and the Weizmann Institute for hospitality,
including the period during which this paper was conceived. SJ
acknowledges partial support from the UK STFC under Grant Agreement
ST/L000504/1, and from the IPPP through an associateship.
SJ acknowledges the NExT Institute.

\vspace{-2mm}

\bibliographystyle{utphys}
\bibliography{references}

\providecommand{\href}[2]{#2}\begingroup\raggedright\begin{thebibliography}{10}

\bibitem{ATLAS-CONF-2015-081}
{ ATLAS} Collaboration, ``{Search for resonances decaying to photon pairs in
  3.2~fb$^{-1}$ of $pp$ collisions at $\sqrt s = 13$~TeV with the ATLAS
  detector},'' Tech. Rep. ATLAS-CONF-2015-081, CERN, Geneva, 2015.
\newblock \url{http://cds.cern.ch/record/2114853}.

\bibitem{CMS-PAS-EXO-15-004}
{ CMS} Collaboration, ``{Search for new physics in high mass diphoton events in
  proton-proton collisions at 13~TeV},'' Tech. Rep. CMS-PAS-EXO-15-004, CERN,
  Geneva, 2015.
\newblock \url{http://cds.cern.ch/record/2114808}.

\bibitem{Landau:1948kw}
L.~D. Landau, ``{On the angular momentum of a system of two photons},''
\href{http://dx.doi.org/10.1016/B978-0-08-010586-4.50070-5}{{\em Dokl. Akad.
  Nauk Ser. Fiz.} {\bfseries 60} (1948) 207}.
%%CITATION = DANKA,60,207;%%.

\bibitem{Yang:1950rg}
C.-N. Yang, ``{Selection Rules for the Dematerialization of a Particle Into Two
  Photons},''
\href{http://dx.doi.org/10.1103/PhysRev.77.242}{{\em Phys. Rev.} {\bfseries 77}
  (1950) 242}.
%%CITATION = PHRVA,77,242;%%.

\bibitem{Alwall:2014hca}
J.~Alwall, R.~Frederix, S.~Frixione, V.~Hirschi, F.~Maltoni, O.~Mattelaer,
  H.-S. Shao, T.~Stelzer, P.~Torrielli, and M.~Zaro, ``{The automated
  computation of tree-level and next-to-leading order differential cross
  sections, and their matching to parton shower simulations},''
  \href{http://dx.doi.org/10.1007/JHEP07(2014)079}{{\em JHEP} {\bfseries 1407}
  (2014) 079},
\href{http://arxiv.org/abs/1405.0301}{{\ttfamily arXiv:1405.0301 [hep-ph]}}.
%%CITATION = ARXIV:1405.0301;%%.

\bibitem{Alloul:2013bka}
A.~Alloul, N.~D. Christensen, C.~Degrande, C.~Duhr, and B.~Fuks, ``{FeynRules
  2.0 - A complete toolbox for tree-level phenomenology},''
  \href{http://dx.doi.org/10.1016/j.cpc.2014.04.012}{{\em Comput. Phys.
  Commun.} {\bfseries 185} (2014) 2250},
\href{http://arxiv.org/abs/1310.1921}{{\ttfamily arXiv:1310.1921 [hep-ph]}}.
%%CITATION = ARXIV:1310.1921;%%.

\bibitem{Dawson:1984gx}
S.~Dawson, ``{The Effective $W$ Approximation},''
\href{http://dx.doi.org/10.1016/0550-3213(85)90038-0}{{\em Nucl. Phys.}
  {\bfseries B249} (1985) 42}.
%%CITATION = NUPHA,B249,42;%%.

\bibitem{Kane:1984bb}
G.~L. Kane, W.~W. Repko, and W.~B. Rolnick, ``{The Effective $W^\pm$, $Z^0$
  Approximation for High-Energy Collisions},''
\href{http://dx.doi.org/10.1016/0370-2693(84)90105-9}{{\em Phys. Lett.}
  {\bfseries B148} (1984) 367}.
%%CITATION = PHLTA,B148,367;%%.

\bibitem{Ball:2012cx}
R.~D. Ball {\em et~al.}, ``{Parton distributions with LHC data},''
  \href{http://dx.doi.org/10.1016/j.nuclphysb.2012.10.003}{{\em Nucl. Phys.}
  {\bfseries B867} (2013) 244},
\href{http://arxiv.org/abs/1207.1303}{{\ttfamily arXiv:1207.1303 [hep-ph]}}.
%%CITATION = ARXIV:1207.1303;%%.

\bibitem{Schmidt:2015zda}
C.~Schmidt, J.~Pumplin, D.~Stump, and C.~P. Yuan, ``{CT14QED parton
  distribution functions from isolated photon production in deep inelastic
  scattering},'' \href{http://dx.doi.org/10.1103/PhysRevD.93.114015}{{\em Phys.
  Rev.} {\bfseries D93} no.~11, (2016) 114015},
\href{http://arxiv.org/abs/1509.02905}{{\ttfamily arXiv:1509.02905 [hep-ph]}}.
%%CITATION = ARXIV:1509.02905;%%.

\bibitem{Harland-Lang:2016qjy}
L.~A. Harland-Lang, V.~A. Khoze, and M.~G. Ryskin, ``{The production of a
  diphoton resonance via photon-photon fusion},''
  \href{http://dx.doi.org/10.1007/JHEP03(2016)182}{{\em JHEP} {\bfseries 03}
  (2016) 182},
\href{http://arxiv.org/abs/1601.07187}{{\ttfamily arXiv:1601.07187 [hep-ph]}}.
%%CITATION = ARXIV:1601.07187;%%.

\bibitem{Altmannshofer:2015xfo}
W.~Altmannshofer, J.~Galloway, S.~Gori, A.~L. Kagan, A.~Martin, and J.~Zupan,
  ``{On the 750~GeV di-photon excess},''
\href{http://arxiv.org/abs/1512.07616}{{\ttfamily arXiv:1512.07616 [hep-ph]}}.
%%CITATION = ARXIV:1512.07616;%%.

\bibitem{Fichet:2015vvy}
S.~Fichet, G.~von Gersdorff, and C.~Royon, ``{Scattering Light by Light at
  750~GeV at the LHC},''
\href{http://arxiv.org/abs/1512.05751}{{\ttfamily arXiv:1512.05751 [hep-ph]}}.
%%CITATION = ARXIV:1512.05751;%%.

\bibitem{CMS:2015neg}
{ CMS} Collaboration, ``{Search for Resonances Decaying to Dijet Final States
  at $\sqrt{s} = 8$~TeV with Scouting Data},'' Tech. Rep. CMS-PAS-EXO-14-005,
  CERN, Geneva, 2015.
\newblock \url{http://cds.cern.ch/record/2063491}.

\bibitem{Khachatryan:2015qba}
{ CMS} Collaboration, V.~Khachatryan {\em et~al.}, ``{Search for diphoton
  resonances in the mass range from 150 to 850 GeV in pp collisions at
  $\sqrt{s} =$ 8 TeV},''
  \href{http://dx.doi.org/10.1016/j.physletb.2015.09.062}{{\em Phys. Lett.}
  {\bfseries B750} (2015) 494},
\href{http://arxiv.org/abs/1506.02301}{{\ttfamily arXiv:1506.02301 [hep-ex]}}.
%%CITATION = ARXIV:1506.02301;%%.

\bibitem{Aad:2015mna}
{ ATLAS} Collaboration, G.~Aad {\em et~al.}, ``{Search for high-mass diphoton
  resonances in $pp$ collisions at $\sqrt{s}=8$ TeV with the ATLAS detector},''
  \href{http://dx.doi.org/10.1103/PhysRevD.92.032004}{{\em Phys. Rev.}
  {\bfseries D92} (2015) 032004},
\href{http://arxiv.org/abs/1504.05511}{{\ttfamily arXiv:1504.05511 [hep-ex]}}.
%%CITATION = ARXIV:1504.05511;%%.

\bibitem{CMS-PAS-EXO-12-045}
{ CMS} Collaboration, ``{Search for High-Mass Diphoton Resonances in pp
  Collisions at $\sqrt s=8$~TeV with the CMS Detector},'' Tech. Rep.
  CMS-PAS-EXO-12-045, CERN, Geneva, 2015.
\newblock \url{http://cds.cern.ch/record/2017806}.

\bibitem{Khachatryan:2015sma}
{ CMS} Collaboration, V.~Khachatryan {\em et~al.}, ``{Search for Resonant
  $\mathrm{t\bar{t}}$ Production in Proton-Proton Collisions at $\sqrt{s}$ = 8
  TeV},''
\href{http://arxiv.org/abs/1506.03062}{{\ttfamily arXiv:1506.03062 [hep-ex]}}.
%%CITATION = ARXIV:1506.03062;%%.

\bibitem{Aad:2015agg}
{ ATLAS} Collaboration, G.~Aad {\em et~al.}, ``{Search for a high-mass Higgs
  boson decaying to a $W$ boson pair in $pp$ collisions at $\sqrt{s} = 8$ TeV
  with the ATLAS detector},''
\href{http://arxiv.org/abs/1509.00389}{{\ttfamily arXiv:1509.00389 [hep-ex]}}.
%%CITATION = ARXIV:1509.00389;%%.

\bibitem{CMS-PAS-HIG-14-007}
{ CMS} Collaboration, ``{Search for a standard model like Higgs boson in the $H
  \to ZZ \to \ell^+\ell^-q\bar q$ decay channel at $\sqrt s=8$~TeV},'' Tech.
  Rep. CMS-PAS-HIG-14-007, CERN, Geneva, 2015.
\newblock \url{https://cds.cern.ch/record/2001558}.

\bibitem{CMS-PAS-HIG-14-013}
{ CMS} Collaboration, ``{Search for di-Higgs resonances decaying to 4 bottom
  quarks},'' Tech. Rep. CMS-PAS-HIG-14-013, CERN, Geneva, 2014.
\newblock \url{http://cds.cern.ch/record/1748425}.

\bibitem{Aad:2015yza}
{ ATLAS} Collaboration, G.~Aad {\em et~al.}, ``{Search for a new resonance
  decaying to a $W$ or $Z$ boson and a Higgs boson in the $\ell \ell / \ell \nu
  / \nu \nu + b \bar{b}$ final states with the ATLAS detector},''
  \href{http://dx.doi.org/10.1140/epjc/s10052-015-3474-x}{{\em Eur. Phys. J.}
  {\bfseries C75} (2015) 263},
\href{http://arxiv.org/abs/1503.08089}{{\ttfamily arXiv:1503.08089 [hep-ex]}}.
%%CITATION = ARXIV:1503.08089;%%.

\bibitem{Aad:2014vgg}
{ ATLAS} Collaboration, G.~Aad {\em et~al.}, ``{Search for neutral Higgs bosons
  of the minimal supersymmetric standard model in $pp$ collisions at $\sqrt{s}
  = 8$~TeV with the ATLAS detector},''
  \href{http://dx.doi.org/10.1007/JHEP11(2014)056}{{\em JHEP} {\bfseries 11}
  (2014) 056},
\href{http://arxiv.org/abs/1409.6064}{{\ttfamily arXiv:1409.6064 [hep-ex]}}.
%%CITATION = ARXIV:1409.6064;%%.

\bibitem{Aad:2014fha}
{ ATLAS} Collaboration, G.~Aad {\em et~al.}, ``{Search for new resonances in
  $W\gamma$ and $Z\gamma$ final states in $pp$ collisions at $\sqrt s=8$ TeV
  with the ATLAS detector},''
  \href{http://dx.doi.org/10.1016/j.physletb.2014.10.002}{{\em Phys. Lett.}
  {\bfseries B738} (2014) 428},
\href{http://arxiv.org/abs/1407.8150}{{\ttfamily arXiv:1407.8150 [hep-ex]}}.
%%CITATION = ARXIV:1407.8150;%%.

\bibitem{Aad:2014cka}
{ ATLAS} Collaboration, G.~Aad {\em et~al.}, ``{Search for high-mass dilepton
  resonances in $pp$ collisions at $\sqrt{s}=8$ TeV with the ATLAS detector},''
  \href{http://dx.doi.org/10.1103/PhysRevD.90.052005}{{\em Phys. Rev.}
  {\bfseries D90} (2014) 052005},
\href{http://arxiv.org/abs/1405.4123}{{\ttfamily arXiv:1405.4123 [hep-ex]}}.
%%CITATION = ARXIV:1405.4123;%%.

\bibitem{CMS:2012yf}
{ CMS} Collaboration, S.~Chatrchyan {\em et~al.}, ``{Search for narrow
  resonances and quantum black holes in inclusive and $b$-tagged dijet mass
  spectra from $pp$ collisions at $\sqrt{s}=7$ TeV},''
  \href{http://dx.doi.org/10.1007/JHEP01(2013)013}{{\em JHEP} {\bfseries 01}
  (2013) 013},
\href{http://arxiv.org/abs/1210.2387}{{\ttfamily arXiv:1210.2387 [hep-ex]}}.
%%CITATION = ARXIV:1210.2387;%%.

\bibitem{CMS-PAS-EXO-12-023}
{ CMS} Collaboration, ``{Search for Heavy Resonances Decaying into $b\bar b$
  and $bg$ Final States in $pp$ Collisions at $\sqrt s = 8$~TeV},'' Tech. Rep.
  CMS-PAS-EXO-12-023, CERN, Geneva, 2013.
\newblock \url{http://cds.cern.ch/record/1542405}.

\bibitem{ATLAS:2015nsi}
{ ATLAS} Collaboration, ``{Search for New Phenomena in Dijet Mass and Angular
  Distributions from $pp$ Collisions at $\sqrt{s}$ = 13 TeV with the ATLAS
  Detector},''
\href{http://arxiv.org/abs/1512.01530}{{\ttfamily arXiv:1512.01530 [hep-ex]}}.
%%CITATION = ARXIV:1512.01530;%%.

\bibitem{Khachatryan:2015dcf}
{ CMS} Collaboration, V.~Khachatryan {\em et~al.}, ``{Search for narrow
  resonances decaying to dijets in proton-proton collisions at $\sqrt s =
  13$~TeV},''
\href{http://arxiv.org/abs/1512.01224}{{\ttfamily arXiv:1512.01224 [hep-ex]}}.
%%CITATION = ARXIV:1512.01224;%%.

\bibitem{Goldberger:2008zz}
W.~D. Goldberger, B.~Grinstein, and W.~Skiba, ``{Distinguishing the Higgs boson
  from the dilaton at the Large Hadron Collider},''
  \href{http://dx.doi.org/10.1103/PhysRevLett.100.111802}{{\em Phys. Rev.
  Lett.} {\bfseries 100} (2008) 111802},
\href{http://arxiv.org/abs/0708.1463}{{\ttfamily arXiv:0708.1463 [hep-ph]}}.
%%CITATION = ARXIV:0708.1463;%%.

\bibitem{Robens:2015gla}
T.~Robens and T.~Stefaniak, ``{Status of the Higgs Singlet Extension of the
  Standard Model after LHC Run 1},''
  \href{http://dx.doi.org/10.1140/epjc/s10052-015-3323-y}{{\em Eur. Phys. J.}
  {\bfseries C75} (2015) 104},
\href{http://arxiv.org/abs/1501.02234}{{\ttfamily arXiv:1501.02234 [hep-ph]}}.
%%CITATION = ARXIV:1501.02234;%%.

\bibitem{Spira:1995rr}
M.~Spira, A.~Djouadi, D.~Graudenz, and P.~M. Zerwas, ``{Higgs boson production
  at the LHC},'' \href{http://dx.doi.org/10.1016/0550-3213(95)00379-7}{{\em
  Nucl. Phys.} {\bfseries B453} (1995) 17},
\href{http://arxiv.org/abs/hep-ph/9504378}{{\ttfamily arXiv:hep-ph/9504378}}.
%%CITATION = HEP-PH/9504378;%%.

\bibitem{Son:2015vfl}
M.~Son and A.~Urbano, ``{A new scalar resonance at 750 GeV: Towards a proof of
  concept in favor of strongly interacting theories},''
\href{http://arxiv.org/abs/1512.08307}{{\ttfamily arXiv:1512.08307 [hep-ph]}}.
%%CITATION = ARXIV:1512.08307;%%.

\bibitem{Xiao:2014kba}
M.-L. Xiao and J.-H. Yu, ``{Stabilizing electroweak vacuum in a vectorlike
  fermion model},'' \href{http://dx.doi.org/10.1103/PhysRevD.90.014007}{{\em
  Phys. Rev.} {\bfseries D90} (2014) 014007},
\href{http://arxiv.org/abs/1404.0681}{{\ttfamily arXiv:1404.0681 [hep-ph]}}.
%%CITATION = ARXIV:1404.0681;%%.

\bibitem{Gripaios:2009pe}
B.~Gripaios, A.~Pomarol, F.~Riva, and J.~Serra, ``{Beyond the Minimal Composite
  Higgs Model},'' \href{http://dx.doi.org/10.1088/1126-6708/2009/04/070}{{\em
  JHEP} {\bfseries 04} (2009) 070},
\href{http://arxiv.org/abs/0902.1483}{{\ttfamily arXiv:0902.1483 [hep-ph]}}.
%%CITATION = ARXIV:0902.1483;%%.

\bibitem{Efrati:2014aea}
A.~Efrati, E.~Kuflik, S.~Nussinov, Y.~Soreq, and T.~Volansky, ``{Constraining
  the Higgs-Dilaton with LHC and Dark Matter Searches},''
  \href{http://dx.doi.org/10.1103/PhysRevD.91.055034}{{\em Phys. Rev.}
  {\bfseries D91} (2015) 055034},
\href{http://arxiv.org/abs/1410.2225}{{\ttfamily arXiv:1410.2225 [hep-ph]}}.
%%CITATION = ARXIV:1410.2225;%%.

\bibitem{Wells:2009kq}
J.~D. Wells, ``{Lectures on Higgs Boson Physics in the Standard Model and
  Beyond},'' in {\em {39th British Universities Summer School in Theoretical
  Elementary Particle Physics (BUSSTEPP 2009) Liverpool, United Kingdom, August
  24-September 4, 2009}}.
\newblock
\href{http://arxiv.org/abs/0909.4541}{{\ttfamily arXiv:0909.4541 [hep-ph]}}.
\newblock
%%CITATION = ARXIV:0909.4541;%%.

\bibitem{Bellazzini:2012vz}
B.~Bellazzini, C.~Csaki, J.~Hubisz, J.~Serra, and J.~Terning, ``{A Higgslike
  Dilaton},'' \href{http://dx.doi.org/10.1140/epjc/s10052-013-2333-x}{{\em Eur.
  Phys. J.} {\bfseries C73} (2013) 2333},
\href{http://arxiv.org/abs/1209.3299}{{\ttfamily arXiv:1209.3299 [hep-ph]}}.
%%CITATION = ARXIV:1209.3299;%%.

\bibitem{Gupta:2009wn}
R.~S. Gupta and J.~D. Wells, ``{Next Generation Higgs Bosons: Theory,
  Constraints and Discovery Prospects at the Large Hadron Collider},''
  \href{http://dx.doi.org/10.1103/PhysRevD.81.055012}{{\em Phys. Rev.}
  {\bfseries D81} (2010) 055012},
\href{http://arxiv.org/abs/0912.0267}{{\ttfamily arXiv:0912.0267 [hep-ph]}}.
%%CITATION = ARXIV:0912.0267;%%.

\bibitem{Gunion:2002zf}
J.~F. Gunion and H.~E. Haber, ``{The CP conserving two Higgs doublet model: The
  Approach to the decoupling limit},''
  \href{http://dx.doi.org/10.1103/PhysRevD.67.075019}{{\em Phys. Rev.}
  {\bfseries D67} (2003) 075019},
\href{http://arxiv.org/abs/hep-ph/0207010}{{\ttfamily arXiv:hep-ph/0207010}}.
%%CITATION = HEP-PH/0207010;%%.

\bibitem{Gupta:2012fy}
R.~S. Gupta, M.~Montull, and F.~Riva, ``{SUSY Faces its Higgs Couplings},''
  \href{http://dx.doi.org/10.1007/JHEP04(2013)132}{{\em JHEP} {\bfseries 04}
  (2013) 132},
\href{http://arxiv.org/abs/1212.5240}{{\ttfamily arXiv:1212.5240 [hep-ph]}}.
%%CITATION = ARXIV:1212.5240;%%.

\bibitem{Branco:2011iw}
G.~C. Branco, P.~M. Ferreira, L.~Lavoura, M.~N. Rebelo, M.~Sher, and J.~P.
  Silva, ``{Theory and phenomenology of two-Higgs-doublet models},''
  \href{http://dx.doi.org/10.1016/j.physrep.2012.02.002}{{\em Phys. Rept.}
  {\bfseries 516} (2012) 1},
\href{http://arxiv.org/abs/1106.0034}{{\ttfamily arXiv:1106.0034 [hep-ph]}}.
%%CITATION = ARXIV:1106.0034;%%.

\bibitem{Ghosh:2015gpa}
D.~Ghosh, R.~S. Gupta, and G.~Perez, ``{Is the Higgs Mechanism of Fermion Mass
  Generation a Fact? A Yukawa-less First-Two-Generation Model},''
\href{http://arxiv.org/abs/1508.01501}{{\ttfamily arXiv:1508.01501 [hep-ph]}}.
%%CITATION = ARXIV:1508.01501;%%.

\bibitem{Drees:1993yr}
M.~Drees and M.~M. Nojiri, ``{Proposed new signal for scalar top-squark
  bound-state production},''
  \href{http://dx.doi.org/10.1103/PhysRevLett.72.2324}{{\em Phys. Rev. Lett.}
  {\bfseries 72} (1994) 2324},
\href{http://arxiv.org/abs/hep-ph/9310209}{{\ttfamily arXiv:hep-ph/9310209}}.
%%CITATION = HEP-PH/9310209;%%.

\bibitem{Kahawala:2011pc}
D.~Kahawala and Y.~Kats, ``{Distinguishing spins at the LHC using bound state
  signals},'' \href{http://dx.doi.org/10.1007/JHEP09(2011)099}{{\em JHEP}
  {\bfseries 09} (2011) 099},
\href{http://arxiv.org/abs/1103.3503}{{\ttfamily arXiv:1103.3503 [hep-ph]}}.
%%CITATION = ARXIV:1103.3503;%%.

\bibitem{Kauth:2009ud}
M.~R. Kauth, J.~H. Kuhn, P.~Marquard, and M.~Steinhauser, ``{Gluinonia: Energy
  Levels, Production and Decay},''
  \href{http://dx.doi.org/10.1016/j.nuclphysb.2010.01.019}{{\em Nucl. Phys.}
  {\bfseries B831} (2010) 285},
\href{http://arxiv.org/abs/0910.2612}{{\ttfamily arXiv:0910.2612 [hep-ph]}}.
%%CITATION = ARXIV:0910.2612;%%.

\bibitem{Frere:1983ag}
J.~M. Frere, D.~R.~T. Jones, and S.~Raby, ``{Fermion Masses and Induction of
  the Weak Scale by Supergravity},''
\href{http://dx.doi.org/10.1016/0550-3213(83)90606-5}{{\em Nucl. Phys.}
  {\bfseries B222} (1983) 11}.
%%CITATION = NUPHA,B222,11;%%.

\bibitem{Derendinger:1983bz}
J.~P. Derendinger and C.~A. Savoy, ``{Quantum Effects and SU(2) x U(1) Breaking
  in Supergravity Gauge Theories},''
\href{http://dx.doi.org/10.1016/0550-3213(84)90162-7}{{\em Nucl. Phys.}
  {\bfseries B237} (1984) 307}.
%%CITATION = NUPHA,B237,307;%%.

\bibitem{Casas:1996de}
J.~A. Casas and S.~Dimopoulos, ``{Stability bounds on flavor violating
  trilinear soft terms in the MSSM},''
  \href{http://dx.doi.org/10.1016/0370-2693(96)01000-3}{{\em Phys. Lett.}
  {\bfseries B387} (1996) 107},
\href{http://arxiv.org/abs/hep-ph/9606237}{{\ttfamily arXiv:hep-ph/9606237}}.
%%CITATION = HEP-PH/9606237;%%.

\bibitem{Rattazzi:1995gk}
R.~Rattazzi and U.~Sarid, ``{The Unified minimal supersymmetric model with
  large Yukawa couplings},''
  \href{http://dx.doi.org/10.1103/PhysRevD.53.1553}{{\em Phys. Rev.} {\bfseries
  D53} (1996) 1553},
\href{http://arxiv.org/abs/hep-ph/9505428}{{\ttfamily arXiv:hep-ph/9505428}}.
%%CITATION = HEP-PH/9505428;%%.

\bibitem{Hisano:2010re}
J.~Hisano and S.~Sugiyama, ``{Charge-breaking constraints on left-right mixing
  of stau's},'' \href{http://dx.doi.org/10.1016/j.physletb.2010.12.013}{{\em
  Phys. Lett.} {\bfseries B696} (2011) 92},
  \href{http://arxiv.org/abs/1011.0260}{{\ttfamily arXiv:1011.0260 [hep-ph]}}.
\href{http://dx.doi.org/10.1016/j.physletb.2013.01.018}{[Erratum: {\em Phys.
  Lett.} {\bf B719} (2013) 472]}.
%%CITATION = ARXIV:1011.0260;%%.

\bibitem{Altmannshofer:2012ks}
W.~Altmannshofer, M.~Carena, N.~R. Shah, and F.~Yu, ``{Indirect Probes of the
  MSSM after the Higgs Discovery},''
  \href{http://dx.doi.org/10.1007/JHEP01(2013)160}{{\em JHEP} {\bfseries 01}
  (2013) 160},
\href{http://arxiv.org/abs/1211.1976}{{\ttfamily arXiv:1211.1976 [hep-ph]}}.
%%CITATION = ARXIV:1211.1976;%%.

\bibitem{Carena:2012mw}
M.~Carena, S.~Gori, I.~Low, N.~R. Shah, and C.~E.~M. Wagner, ``{Vacuum
  Stability and Higgs Diphoton Decays in the MSSM},''
  \href{http://dx.doi.org/10.1007/JHEP02(2013)114}{{\em JHEP} {\bfseries 02}
  (2013) 114},
\href{http://arxiv.org/abs/1211.6136}{{\ttfamily arXiv:1211.6136 [hep-ph]}}.
%%CITATION = ARXIV:1211.6136;%%.

\bibitem{Altmannshofer:2014qha}
W.~Altmannshofer, C.~Frugiuele, and R.~Harnik, ``{Fermion Hierarchy from
  Sfermion Anarchy},'' \href{http://dx.doi.org/10.1007/JHEP12(2014)180}{{\em
  JHEP} {\bfseries 12} (2014) 180},
\href{http://arxiv.org/abs/1409.2522}{{\ttfamily arXiv:1409.2522 [hep-ph]}}.
%%CITATION = ARXIV:1409.2522;%%.

\bibitem{Djouadi:2005gj}
A.~Djouadi, ``{The Anatomy of electro-weak symmetry breaking. II. The Higgs
  bosons in the minimal supersymmetric model},''
  \href{http://dx.doi.org/10.1016/j.physrep.2007.10.005}{{\em Phys. Rept.}
  {\bfseries 459} (2008) 1},
\href{http://arxiv.org/abs/hep-ph/0503173}{{\ttfamily arXiv:hep-ph/0503173}}.
%%CITATION = HEP-PH/0503173;%%.

\bibitem{Borzumati:1999sp}
F.~Borzumati, G.~R. Farrar, N.~Polonsky, and S.~D. Thomas, ``{Soft Yukawa
  couplings in supersymmetric theories},''
  \href{http://dx.doi.org/10.1016/S0550-3213(99)00328-4}{{\em Nucl. Phys.}
  {\bfseries B555} (1999) 53},
\href{http://arxiv.org/abs/hep-ph/9902443}{{\ttfamily arXiv:hep-ph/9902443}}.
%%CITATION = HEP-PH/9902443;%%.

\bibitem{Harigaya:2015ezk}
K.~Harigaya and Y.~Nomura, ``{Composite Models for the 750~GeV Diphoton
  Excess},''
\href{http://arxiv.org/abs/1512.04850}{{\ttfamily arXiv:1512.04850 [hep-ph]}}.
%%CITATION = ARXIV:1512.04850;%%.

\bibitem{Mambrini:2015wyu}
Y.~Mambrini, G.~Arcadi, and A.~Djouadi, ``{The LHC diphoton resonance and dark
  matter},''
\href{http://arxiv.org/abs/1512.04913}{{\ttfamily arXiv:1512.04913 [hep-ph]}}.
%%CITATION = ARXIV:1512.04913;%%.

\bibitem{Backovic:2015fnp}
M.~Backovic, A.~Mariotti, and D.~Redigolo, ``{Di-photon excess illuminates Dark
  Matter},''
\href{http://arxiv.org/abs/1512.04917}{{\ttfamily arXiv:1512.04917 [hep-ph]}}.
%%CITATION = ARXIV:1512.04917;%%.

\bibitem{Angelescu:2015uiz}
A.~Angelescu, A.~Djouadi, and G.~Moreau, ``{Scenarii for interpretations of the
  LHC diphoton excess: two Higgs doublets and vector-like quarks and
  leptons},''
\href{http://arxiv.org/abs/1512.04921}{{\ttfamily arXiv:1512.04921 [hep-ph]}}.
%%CITATION = ARXIV:1512.04921;%%.

\bibitem{Nakai:2015ptz}
Y.~Nakai, R.~Sato, and K.~Tobioka, ``{Footprints of New Strong Dynamics via
  Anomaly},''
\href{http://arxiv.org/abs/1512.04924}{{\ttfamily arXiv:1512.04924 [hep-ph]}}.
%%CITATION = ARXIV:1512.04924;%%.

\bibitem{Knapen:2015dap}
S.~Knapen, T.~Melia, M.~Papucci, and K.~Zurek, ``{Rays of light from the
  LHC},''
\href{http://arxiv.org/abs/1512.04928}{{\ttfamily arXiv:1512.04928 [hep-ph]}}.
%%CITATION = ARXIV:1512.04928;%%.

\bibitem{Buttazzo:2015txu}
D.~Buttazzo, A.~Greljo, and D.~Marzocca, ``{Knocking on New Physics' door with
  a Scalar Resonance},''
\href{http://arxiv.org/abs/1512.04929}{{\ttfamily arXiv:1512.04929 [hep-ph]}}.
%%CITATION = ARXIV:1512.04929;%%.

\bibitem{Pilaftsis:2015ycr}
A.~Pilaftsis, ``{Diphoton Signatures from Heavy Axion Decays at LHC},''
\href{http://arxiv.org/abs/1512.04931}{{\ttfamily arXiv:1512.04931 [hep-ph]}}.
%%CITATION = ARXIV:1512.04931;%%.

\bibitem{Franceschini:2015kwy}
R.~Franceschini, G.~F. Giudice, J.~F. Kamenik, M.~McCullough, A.~Pomarol,
  R.~Rattazzi, M.~Redi, F.~Riva, A.~Strumia, and R.~Torre, ``{What is the
  $\gamma\gamma$ resonance at 750~GeV?},''
\href{http://arxiv.org/abs/1512.04933}{{\ttfamily arXiv:1512.04933 [hep-ph]}}.
%%CITATION = ARXIV:1512.04933;%%.

\bibitem{DiChiara:2015vdm}
S.~Di~Chiara, L.~Marzola, and M.~Raidal, ``{First interpretation of the 750~GeV
  di-photon resonance at the LHC},''
\href{http://arxiv.org/abs/1512.04939}{{\ttfamily arXiv:1512.04939 [hep-ph]}}.
%%CITATION = ARXIV:1512.04939;%%.

\bibitem{Higaki:2015jag}
T.~Higaki, K.~S. Jeong, N.~Kitajima, and F.~Takahashi, ``{The QCD Axion from
  Aligned Axions and Diphoton Excess},''
\href{http://arxiv.org/abs/1512.05295}{{\ttfamily arXiv:1512.05295 [hep-ph]}}.
%%CITATION = ARXIV:1512.05295;%%.

\bibitem{McDermott:2015sck}
S.~D. McDermott, P.~Meade, and H.~Ramani, ``{Singlet Scalar Resonances and the
  Diphoton Excess},''
\href{http://arxiv.org/abs/1512.05326}{{\ttfamily arXiv:1512.05326 [hep-ph]}}.
%%CITATION = ARXIV:1512.05326;%%.

\bibitem{Ellis:2015oso}
J.~Ellis, S.~A.~R. Ellis, J.~Quevillon, V.~Sanz, and T.~You, ``{On the
  Interpretation of a Possible $\sim 750$~GeV Particle Decaying into $\gamma
  \gamma$},''
\href{http://arxiv.org/abs/1512.05327}{{\ttfamily arXiv:1512.05327 [hep-ph]}}.
%%CITATION = ARXIV:1512.05327;%%.

\bibitem{Low:2015qep}
M.~Low, A.~Tesi, and L.-T. Wang, ``{A pseudoscalar decaying to photon pairs in
  the early LHC run~2 data},''
\href{http://arxiv.org/abs/1512.05328}{{\ttfamily arXiv:1512.05328 [hep-ph]}}.
%%CITATION = ARXIV:1512.05328;%%.

\bibitem{Bellazzini:2015nxw}
B.~Bellazzini, R.~Franceschini, F.~Sala, and J.~Serra, ``{Goldstones in
  Diphotons},''
\href{http://arxiv.org/abs/1512.05330}{{\ttfamily arXiv:1512.05330 [hep-ph]}}.
%%CITATION = ARXIV:1512.05330;%%.

\bibitem{Petersson:2015mkr}
C.~Petersson and R.~Torre, ``{The 750~GeV diphoton excess from the goldstino
  superpartner},''
\href{http://arxiv.org/abs/1512.05333}{{\ttfamily arXiv:1512.05333 [hep-ph]}}.
%%CITATION = ARXIV:1512.05333;%%.

\bibitem{Molinaro:2015cwg}
E.~Molinaro, F.~Sannino, and N.~Vignaroli, ``{Strong dynamics or axion origin
  of the diphoton excess},''
\href{http://arxiv.org/abs/1512.05334}{{\ttfamily arXiv:1512.05334 [hep-ph]}}.
%%CITATION = ARXIV:1512.05334;%%.

\bibitem{Jaeckel:2012yz}
J.~Jaeckel, M.~Jankowiak, and M.~Spannowsky, ``{LHC probes the hidden
  sector},'' \href{http://dx.doi.org/10.1016/j.dark.2013.06.001}{{\em Phys.
  Dark Univ.} {\bfseries 2} (2013) 111},
\href{http://arxiv.org/abs/1212.3620}{{\ttfamily arXiv:1212.3620 [hep-ph]}}.
%%CITATION = ARXIV:1212.3620;%%.

\end{thebibliography}\endgroup

\end{document}